\documentclass[manuscript, sigconf, screen]{acmart}

\usepackage{subcaption}
\usepackage{multirow}
\usepackage{array}
\usepackage{enumitem}

\AtBeginDocument{%
  }

\copyrightyear{2026}
\acmYear{2026}
\setcopyright{cc}
\setcctype{by}
\acmConference[DIS '26]{Designing Interactive Systems Conference}{June 13--17, 2026}{Singapore, Singapore}
\acmBooktitle{Designing Interactive Systems Conference (DIS '26), June 13--17, 2026, Singapore, Singapore}
\acmDOI{10.1145/3800645.3812928}
\acmISBN{979-8-4007-2563-0/2026/06}

\begin{document}

\title{Building A Civic Tool for Community-Police Engagement to Adapt Neighborhood Policing}

\author{Ravinithesh Annapureddy}
\orcid{0000-0003-4383-020X}
\affiliation{%
  \institution{Idiap Research Institute}
  \city{Martigny}
  \country{Switzerland}}
\affiliation{%
  \institution{EPFL}
  \city{Lausanne}
  \country{Switzerland}}
\email{ravinithesh.annapureddy@idiap.ch}

\author{Staņislavs Šeiko}
\orcid{0009-0006-3939-3069}
\affiliation{%
  \institution{Riga Municipal Police}
  \city{Riga}
  \country{Latvia}}
\affiliation{%
  \institution{University of Latvia}
  \city{Riga}
  \country{Latvia}}
\email{stanislavs.seiko@riga.lv}

\author{Natalie Higham-James}
\orcid{0000-0001-5711-0209}
\affiliation{%
  \institution{University of York}
  \city{York}
  \country{United Kingdom}
}
\email{natalie.james@york.ac.uk}

\author{William Droz}
\orcid{0000-0003-0379-2018}
\affiliation{%
  \institution{Idiap Research Institute}
  \city{Martigny}
  \country{Switzerland}
}
\email{william.droz@idiap.ch}

\author{Alessandro Fornaroli}
\orcid{0000-0002-4231-8231}
\affiliation{%
  \institution{Idiap Research Institute}
  \city{Martigny}
  \country{Switzerland}
}
\email{alessandro.fornaroli@alumni.epfl.ch}
\authornote{Former affiliation. The work was conducted while affiliated with the Idiap Research Institute.}

\author{Sarah Vollmer}
\orcid{0009-0009-7889-1855}
\affiliation{%
  \institution{Camino}
  \city{Berlin}
  \country{Germany}}
\email{sarahvollmer@camino-werkstatt.de}


\author{Britta Elena Hecking}
\orcid{0009-0009-8071-8947}
\affiliation{%
  \institution{Camino}
  \city{Berlin}
  \country{Germany}}
\email{brittahecking@camino-werkstatt.de}

\author{Daniel Gatica-Perez}
\orcid{0000-0001-5488-2182}
\affiliation{%
  \institution{Idiap Research Institute}
  \city{Martigny}
  \country{Switzerland}}
\affiliation{%
  \institution{EPFL}
  \city{Lausanne}
  \country{Switzerland}}
\email{gatica@idiap.ch}

\renewcommand{\shortauthors}{Annapureddy et al.}

\begin{abstract}

Data-driven policing often prioritizes incident records over residents’ lived experiences. In the Baltic city of Riga, with a history of distrust and limited community-police engagement, this can further alienate the public. To bridge this gap, we propose a Research through Design (RtD) inquiry into the development of \textit{Par dro\v{s}u R\={\i}gu}, a civic tool for community-data-integrated policing. With municipal police, NGOs, and city staff, we ask how RtD enables stakeholder negotiation and which interaction qualities support trust and the use of combined community and incident data. The co-design process included workshops that surfaced divergent notions of safety; material probes designed as boundary objects to negotiate among stakeholders; and a pilot deployment showing how combining quantitative and qualitative data reshapes engagement and trust. Mixed-methods evaluation suggests increased officer-citizen interaction, but frictions in sustaining stakeholder collaboration. We contribute (i) an empirical RtD inquiry with public institutions, (ii) an artifact combining physical and dashboard interactions, and (iii) reflections on interaction design as a boundary-spanning practice for trust and infrastructuring.

\end{abstract}




\maketitle

\section{Introduction}\label{sec:intro}

Urban public space safety is shaped not only by crime statistics but also by perceptions, experiences, and trust. Police have adopted Decision Support Systems (DSS)~\cite{Sherman2013} and technology-first solutions like surveillance and predictive policing to address safety challenges~\cite{Meijer2019}. However, if these systems neglect socio-cultural context, they can alienate communities and erode trust needed to improve public safety~\cite{brayne2021predict, Erete2015}. Currently, many DSS struggle to incorporate qualitative citizen feedback~\cite{Masso2024} alongside quantitative indicators, creating a gap between how safety is measured institutionally and how it is experienced by the public. In response, community policing has gained prominence~\cite{Skogan2008}, emphasizing collaboration between police and communities. However, its effectiveness depends on high-quality information exchange~\cite{Crawford2023} and is often hindered by unequal power relations and a lack of infrastructure to include citizen perspectives~\cite{Skogan2006, Dantec2013}.

\begin{figure*}[ht]
  \centering
  \includegraphics[width=1.0\linewidth]{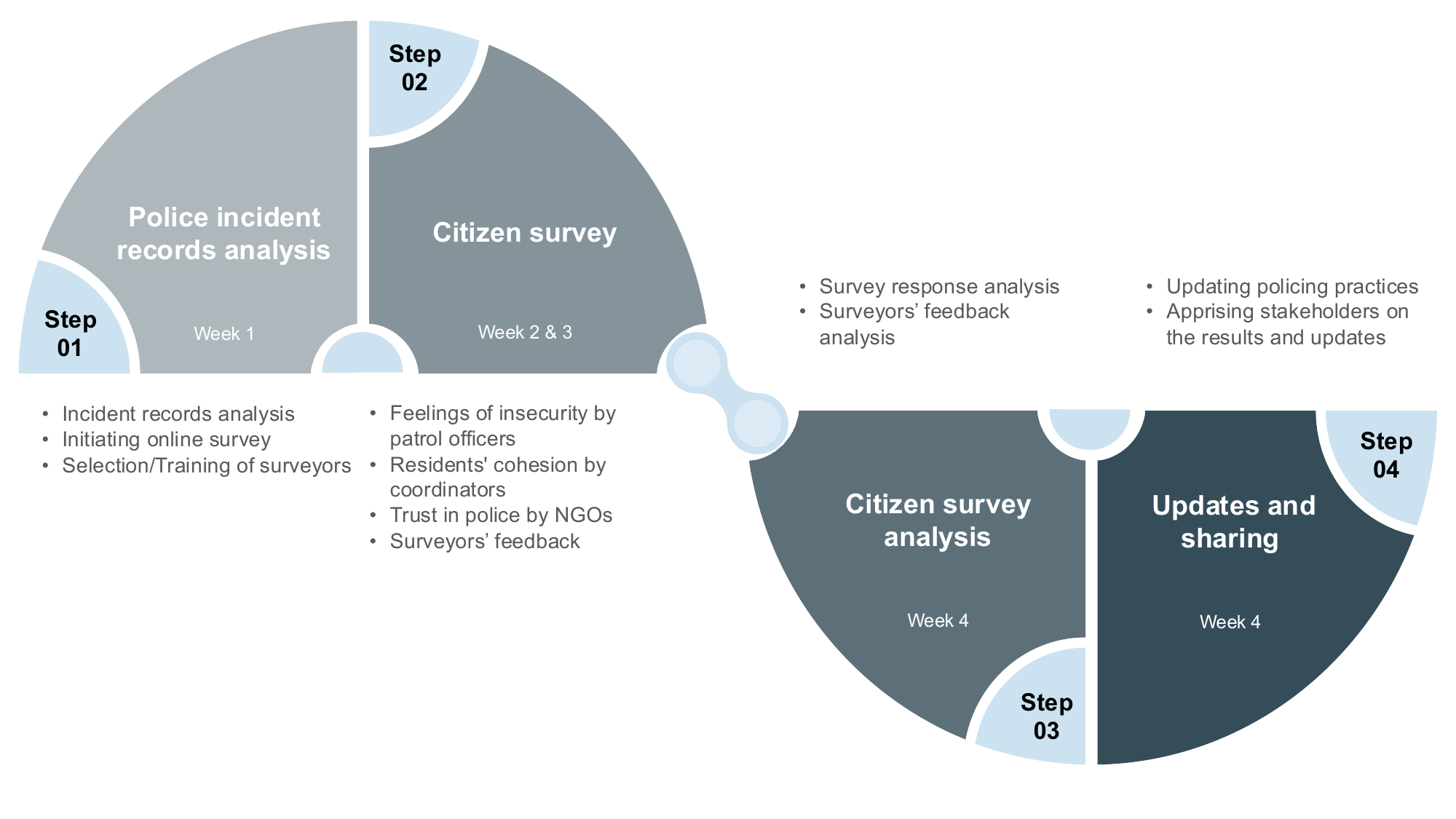}
  \caption{Overview of the \textit{Par dro\v{s}u R\={\i}gu} civic tool with Timeline}
  \Description{Diagram of the process map for the delivery plan of the civic tool. This includes four steps. First, the incident analysis and survey preparation, happening in the first week, includes the incident record analysis, the creation of surveys, the selection of surveyors, and the training of surveyors. Second, the citizen survey, taking the second and third weeks, with the three surveys conducted by the respective surveyors, and the collection of surveyors' feedback. Third, the survey response analysis, conducted by the CoD, happened in the fourth week. And finally, the updates and sharing, with the update of policing practices and appraisal of stakeholders.}
  \label{fig:process-map}
\end{figure*}

Because these challenges are rooted in a lack of trust, they cannot be resolved by top-down technology alone. As an alternative, there is support for human-centered approaches that prioritize relationship building and using technology design as an interaction support~\cite{Davey2005, bradford2010trust}. Consequently, co-design~\cite{Sanders2008, Bødker2022a} has been adopted in civic settings to embed technology within local contexts, enabling communities and authorities to jointly identify needs and create solutions~\cite{Schaap2021, Merkel2004, Manuel2020}. However, applying these bottom-up principles within a top-down hierarchical institution like the police creates frictions~\cite{Dantec2013, Crivellaro2019}.

\begin{sloppypar}
In this paper, we conduct a Research through Design (RtD) inquiry into the frictions of collaborating with police, NGOs, and municipal offices for co-designing a civic tool. We collaborated with these stakeholders in the Baltic city of Riga to co-design the \textit{Par dro\v{s}u R\={\i}gu} civic tool, which supports community-data-integrated policing by integrating and visualizing police incident data alongside citizen perceptions. We define a \textit{Civic Tool} as a socio-technical infrastructure that encourages interaction and engagement between communities and local government, incorporating physical, digital, and human elements. In contrast to tech-centric approaches~\cite{Saldivar2019}, civic tools emphasize human interaction and community-centered design. We selected RtD~\cite{Zimmerman2007} given the wicked nature of the problem, which cannot be fully understood through observation alone; it requires material interventions to reveal underlying tensions. Complementing RtD with co-design lets stakeholders treat emerging artifacts as boundary objects~\cite{Dalsgaard2014}, using them to negotiate and build trust~\cite{Dantec2012, Saldivar2019, Dickinson2021}. In doing so, we draw focus not only to the civic tool itself, but also on how RtD can be conducted as community-engaged research with multiple civic actors involved in policing. 

Co-designing approaches to developing civic tools for community-police engagement remains a relatively unexplored area, particularly in the Baltic region where our research takes place and where community-police engagement is limited. While civic technologies are well established in HCI, RtD with policing institutions is still rare. More importantly, existing RtD accounts have given limited attention to how design operates in contexts where authority, legitimacy, and institutional hierarchy shape what forms of participation and negotiation are possible. This gap is important not only for design research but also in broader discussions of HCI for civic technologies. Our study, therefore, extends RtD in the public sector, specifically in Riga's policing context, by addressing the following research questions:
\end{sloppypar}
\begin{enumerate}
    \item \textbf{RQ1}: How can RtD facilitate negotiation between multiple stakeholders to co-design a civic tool for community-police engagement?
    \item \textbf{RQ2}: What interaction practices support the materialization of community-data-integrated policing and trust relations in Riga's context?
\end{enumerate}

We address these questions through the design and demonstration of the \textit{Par dro\v{s}u R\={\i}gu} civic tool, a neighborhood-level interaction mechanism that integrates police records with situated safety perceptions gathered via an in-person citizen survey to inform neighborhood policing decisions. To provide an anchor for the design process in the paper, we present an overview of the \textit{Par dro\v{s}u R\={\i}gu} civic tool here. Figure~\ref{fig:process-map} shows an overview of its functioning through four steps: (1) analyzing recent incident records of the neighborhood; (2) gathering community perception of safety in a neighborhood by three surveyor groups: patrol officers on citizens' feelings of insecurity, district municipal coordinator on residents' cohesion, and NGOs on residents' trust in the police; (3) analyzing and juxtaposing survey responses with incident records; and (4) using the comparisons as decision support for adjustments to policing practices (e.g., reallocating patrols, convening community meetings). The tool facilitates situated interaction between surveyors engaging residents on the street to gather feedback; human-data interaction of police with a web application to visualize and interpret the data; and institutional interaction between the stakeholders for implementation. We make three contributions:

\begin{sloppypar}
\begin{enumerate}
    \item \textbf{Empirical (RQ1):} A two-year RtD account of a dozen activities showing how co-designing with police, NGOs, and municipal offices surfaces frictions (data literacy, role asymmetry, time constraints) and how material and interactional design probes mediate negotiation (three-part survey, text boxes for visualizations, and badges for survey respondents), offering an infrastructuring model for designing civic technologies. This contribution highlights how RtD can operate in policing as a form of community-engaged research. This is presented in \emph{Section~\ref{sec:methodology}}.

    \item \textbf{Artifactual (RQ1):} The \textit{Par dro\v{s}u R\={\i}gu} civic tool to support community-data-integrated policing with three types of interactions to bridge the gap between police data and citizen perceptions of safety, where surveys are interactional moments while being data collection instruments. This is presented in \emph{Section~\ref{sec:tool}}.

    \item \textbf{Reflective (RQ2):} Our mixed-methods evaluation highlights that the civic tool increased resident-officer interaction and was perceived by city leaders as a mechanism to build trust, while also exposing limits of participation within hierarchical institutions. With these insights, we argue that civic tools must be designed as boundary objects for negotiation and trust-building, not just aiming for usability. This is presented in \emph{Sections~\ref{sec:evaluation} and~\ref{sec:discussion}}.
\end{enumerate}
\end{sloppypar}

\section{Related Work} \label{sec:related-work}

\subsection{RtD with Public Institutions}

Research through Design (RtD)~\cite{Zimmerman2007} is an interaction design research approach well-suited to context-dependent, wicked problems~\cite{Zimmerman2014, Gaver2012}. In these contexts, RtD often functions as a mode of infrastructuring: aligning socio-technical resources, practices, and institutional logics over time. Dalsgaard et al.~\cite{Dalsgaard2014} frame design artifacts as boundary objects that enable cross-disciplinary negotiation; we use this idea to link police records with residents' lived experience. In our case, improving community-police engagement required negotiation beyond technical innovation, making RtD an appropriate approach. Work on co-designing in public institutions similarly emphasizes that outcomes depend not only on creating tools, but on the ongoing maintenance of relationships and other forms of ``human work''~\cite{Bødker2022a}. Designing for the publics can create institutional frictions while also opening pluralistic spaces where differences can be debated~\cite{Crivellaro2019, Dantec2013}. Only a few studies embed RtD within hierarchical public institutions such as policing~\cite{Kirsi2011, Annapureddy2025, Signori2023}. For example, ``Sbocciamo Torino'' used co-design workshops to create a multi-stakeholder collaboration and data analysis system for managing youth deviance~\cite{Annapureddy2025}, showing how the materialization of data workflows helped police officers and social workers negotiate conflicting professional cultures. Another work applied a human-centered design approach to develop a handover protocol to address staff turnover in community policing (the ``GMP Community Connect tool'')~\cite{Signori2023}, highlighting that systems prioritized informational continuity over relational continuity with the community. Critical RtD studies also speculate on ethical implications: one work employed design fiction to critique surveillance in unemployment services, showing the digital labor imposed by public institutions~\cite{Moller2021}. Our work extends this by exploring how RtD can function inside a police institution.

\subsection{Co-Design for Public Safety}

Public spaces can promote social cohesion and inclusion when they are accessible and welcoming to all~\cite{Crawford2023}. Safety challenges can undermine these benefits and reduce quality of life~\cite{Crawford2023}. Because safety is shaped by sociocultural factors and varies across demographic groups, local authorities need tailored approaches that involve communities and support collaborative strategies~\cite{Crawford2023}. In policing, where power asymmetries between law enforcement and citizens are common, co-design can help bridge these divides~\cite{Davey2005}. Co-designing, defined as the creativity of designers and people not trained in design working together in the design development process~\cite{Sanders2008}, has been applied to public safety to build proactive measures for crime and security challenges~\cite{Cozens2015}. Mechanisms such as community meetings, social media platforms, and surveys are commonly used to support this engagement~\cite{Skoric2016}. Work on policing technologies has argued for a shift from one-directional information provision toward collaborative analysis of evidence, including efforts to build trust with digital systems~\cite{Brush2013, Zhang2020}. Co-design with citizens and police can improve communication, enhance trust~\cite{Annapureddy2025, Signori2023}, and produce systems that better reflect community needs by surfacing feedback on social relevance, ethical implications, and practical utility~\cite{Haque2024}. Yet integrating co-designed outputs into official workflows remains difficult. Our study documents the negotiations required to align the open-ended nature of co-design with the operational constraints of the police.

\subsection{Community-Data-Integrated Policing}
\begin{sloppypar}
Decision Support Systems (DSS) are used in policing to predict crime, optimize officer deployment, identify crime hotspots, and guide the practice and evaluation of agencies, units, and officers~\cite{Sherman2013, Meijer2019}. Although DSS can serve as boundary objects to support dialogue between communities and police, they often rely on incident data (e.g., crime statistics, surveillance footage) and do not address nuanced community experiences. To involve communities and capture safety perceptions, we use citizen-generated data (CGD): voluntary reports from residents that provide firsthand insights into their experiences, serving as an alternative to institutional data~\cite{ponti2020citizen}. Previous research has examined platforms for citizens to report issues ranging from urban maintenance (e.g., 311 systems), traffic congestion, and public-space quality to personal safety~\cite{Ming2021, Johnson2018, ponti2020citizen, Alessandro2023UrbanCrowd}. A crowdsourced system for reporting local crimes surfaced incidents that would otherwise go unreported, creating an information repository relied on by community members and some authorities~\cite{Alvarado2018}. Studies also indicate that CGD becomes actionable only when supported by human infrastructure (moderators and data intermediaries) that integrate it into local contexts. Designing civic tools thus involves not just opening data streams but enabling ongoing infrastructuring activities---workshops, curation practices, feedback loops---to ensure citizen feedback informs public services~\cite{Monteiro2013, Peer2023}.
\end{sloppypar}

Surveys are a common method for collecting public safety information. To align policing with community needs, it is essential to understand feelings of insecurity, perceived risk, trust in police, and social cohesion~\cite{Lynch2006}. Community surveys address this gap~\cite{Skogan2014} and are used to identify hotspots~\cite{Haberman2016} and evaluate police performance~\cite{Rosenbaum2017}. These measures can guide the allocation of resources to address specific community needs and concerns~\cite{Tseloni2008FearofCrime}. In our work, we use community surveys as a form of CGD to gather citizens' safety perceptions.

\subsection{Interaction Design for Trust}

Technology-mediated interaction has been positioned as a way to operationalize community policing by promoting citizen engagement with police~\cite{Brewster2018}. Yet because community policing prioritizes trust and collaboration over enforcement~\cite{Skogan2008}, and because trust is enacted through specific interactions~\cite{Skogan2006, Schaap2021}, such systems must address trust, privacy, and motivation for engagement~\cite{Skogan2006, Corbett2018, Corbett2021}. Prior work examines how technology shapes community participation~\cite{Erete2014} and supports involvement in local governance~\cite{Erete2017Empowered}. Studies in complex civic settings show that platforms can broaden participation and increase stakeholder visibility~\cite{Sistrunk2024}, while research on online neighborhood crime discussions suggests that digital systems complement offline community relations rather than replace in-person participation that supports collective action~\cite{Erete2015}. Other work explores how technology mediates relationships between residents and public authority groups~\cite{Erete2013, Dow2018, Crivellaro2019}. Related systems also emphasize institutional transparency, such as police sharing crime statistics through online maps to invite public scrutiny and engagement~\cite{Puussaar2018, Zhang2020}. 

\begin{sloppypar}
Prior systems illustrate both opportunities and tensions in participatory safety technologies. ``Digital Neighborhood Watch'' explores sharing home security camera feeds among neighbors to support situational awareness and coordination, while surfacing concerns about privacy and interpersonal trust~\cite{Brush2013}. ``CampusWatch'' investigates a crowdsourced patrolling model in which volunteers self-schedule patrols and record incidents, highlighting how motivators shape participation and offering implications for cooperative patrolling systems~\cite{Park2018}. Across these settings, stakeholders often interact with the same system in different ways, reflecting diverse roles, expertise, and values~\cite{Haque2024}. A recurring challenge is disagreement over what counts as meaningful information: residents emphasize lived experiences and perceptions of safety, while institutions often privilege quantitative metrics and formalized procedures~\cite{Manuel2020}. This tension is particularly relevant in policing, where official crime data often fails to capture citizens' day-to-day security concerns. At the same time, these examples foreground a core tradeoff: technology can enable participation in safety practices while also introducing new frictions.
\end{sloppypar}

\subsection{Gap}

\begin{sloppypar}
While prior work has explored citizen-police interactions, community-generated safety data, and co-design in civic contexts, most studies focus on a single stakeholder group (citizens or authorities). Few studies have co-designed a civic tool with multiple actors---citizens, police officers, NGOs, and municipal institutions---for public space safety. In this sense, the gap is not only empirical---a lack of policing cases---but also conceptual: the policing context in which institutional hierarchy, authority, and legitimacy play a defining role. We work at this gap by contributing a real-world case of a system that integrates citizen feedback into police workflows through a civic tool. We explored RtD with police and civic actors involved in policing and showed how the policing context sharpens broader RtD questions around asymmetry, interpretation, and infrastructural compatibility between various actors.
\end{sloppypar}

\section{RtD Inquiry} \label{sec:methodology}

\subsection{Public Safety Context in Riga}

Our study takes place in Riga, which is composed of six districts subdivided into 58 neighborhoods. Each district has a municipal office and a police station led by a Police Chief of District (CoD). Local public safety work primarily concerns public order, minor offenses, and neighborhood disputes rather than terrorism or violent organized crime~\cite{Lukass2019}. During preparatory consultations (Section~\ref{sec:preparatory-workshop}), police officers noted that official incident statistics often do not capture residents' lived experiences of insecurity\footnote{Contextual details regarding local police operations rely on the situated institutional knowledge of the second author, who works in the Riga Municipal Police.} or levels of trust in the police~\cite{Barvika2015}. They also described existing feedback mechanisms as limited: the city runs a biennial, multi-topic survey in which safety is covered by only a few questions, and responses are geocoded only at the city level (not by district or neighborhood). Prior work suggests that resident insecurity perceptions and police distrust are shaped by historical and socio-cultural factors, including Soviet-era legacies of mistrust in public institutions and the ethnic and linguistic diversity of neighborhoods~\cite{jurkynas2014}. Together, these factors complicate efforts to build cooperative relationships between citizens and the police. Despite this context, the city's municipal police expressed interest in incorporating citizen perspectives to strengthen mutual relations and trust.

\begin{sloppypar}
Perceived police legitimacy varies substantially across (and within) countries and influences citizens' willingness to engage with the police~\cite{Skogan2006}. Policing attitudes and engagement practices, therefore, differ across contexts. In the United States, community-engagement initiatives can face skepticism linked to systemic discrimination, police militarization, high-profile incidents, and contested accountability~\cite{Israni2017, Narayanan2024}; post-9/11 counterterrorism shifts have also been associated with trust erosion in some communities~\cite{Braga2015}. In parts of Europe, community-police relations are often described as more collaborative or pragmatic, though these dynamics remain uneven and contested across settings~\cite{Eliasson2023, Charlton2023, BBCNews2024, Schaap2021}. Nordic countries are frequently characterized as closer to trust-based, low-conflict models emphasizing mediation and community participation~\cite{Eliasson2023}. Riga has neither extreme distrust nor strong collaboration~\cite{Masso2024}. This motivated us to design a civic tool aimed at bridging institutional incident data and residents' lived experiences. We adopted an RtD approach to co-design the system in a way that accounts for, and builds on, existing community-police relationships~\cite{Saldivar2019}. However, we recognize the global variations and do not intend to generalize our findings universally but rather highlight how local context and co-design approaches can shape policing.
\end{sloppypar}

\subsection{Co-Designing} \label{sec:design-steps}

Over two years, we facilitated a dozen structured activities~\cite{Bødker2022b} with the city's municipal police, city staff, and local NGOs. While the work moved broadly from problem framing to deployment, it remained iterative across four phases: (1) Framing, Alignment, and Requirements; (2) Prototyping and Iteration; (3) Validation and Demonstration; and (4) Assessment. This section summarizes the activities; additional details are provided in Appendix~\ref{app:pd-process}.

\subsubsection{Framing the Problem, Aligning Stakeholders, and Gathering the Requirements} \label{sec:preparatory-workshop}

\begin{figure}[ht]
  \centering
  \includegraphics[width=1.0\linewidth]{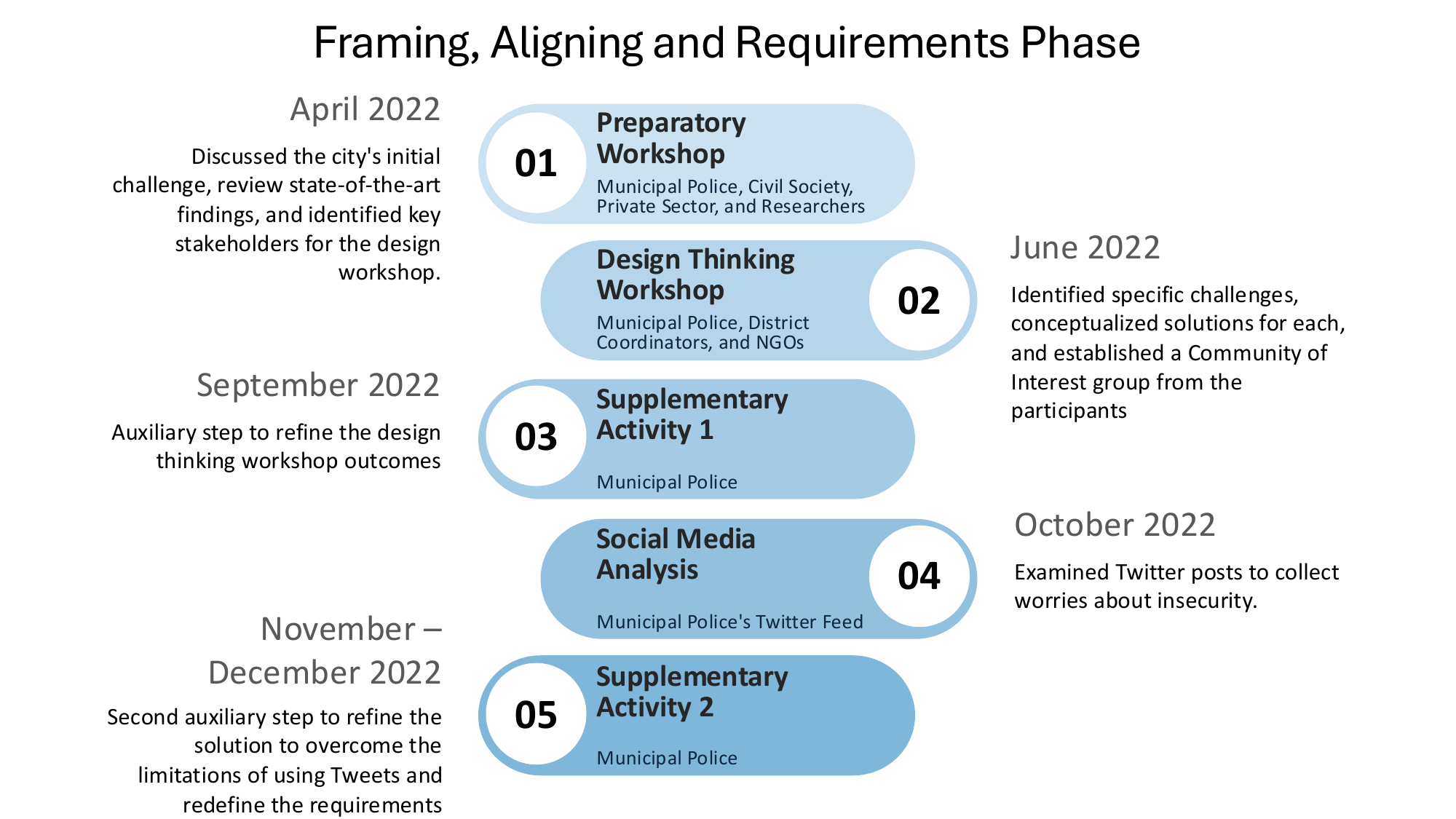}
  \caption{Activities of framing, alignment, and requirements phase}
  \Description{Timeline diagram of the ``Framing, Alignment, and Requirements'' phase (April–December 2022) showing five activities: (01) Preparatory Workshop with municipal police, civil society, private sector, and researchers (April 2022); (02) Design Thinking Workshop with municipal police, district coordinators, and NGOs (June 2022); (03) Supplementary Activity 1 with municipal police (September 2022); (04) Social Media Analysis of the municipal police Twitter feed (October 2022); and (05) Supplementary Activity 2 with municipal police (November–December 2022).}
  \label{fig:pd1}
\end{figure}

The first phase (Figure~\ref{fig:pd1}) began with a \textit{preparatory workshop} \textbf{(Activity 01)} involving representatives from the city's police, academics, and practitioners in HCI, urban security, and public safety. Police highlighted spatial and temporal limitations in existing citizen feedback mechanisms. The activity framed the design challenge as: \textit{In what ways might we gather and understand data to assess risks in public spaces and neighborhoods to improve community-data-integrated policing?} We also identified the stakeholder groups to be involved in the subsequent process.

We then convened a \textit{design thinking workshop} \textbf{(Activity 02)} to identify specific problems and brainstorm potential solutions. Participants included municipal staff from the urban development department, municipal police, an association of restaurants, and four NGOs working on pedestrian and bicyclist rights and environmental improvements in the city. The four groups focused on overly specific everyday problems in public spaces, such as the safety of pedestrians and on streets caused by illegal parking, alcohol, and drug abuse in public spaces, as well as citizens' feelings of insecurity. Such diversity of perceived challenges did not provide a clear direction for solutions. Because the workshop did not converge on a clear solution direction, we conducted a \textit{supplementary activity} \textbf{(Activity 03)} with the police to further probe solutions.

\begin{sloppypar}
Early attempts to pursue purely digital approaches did not yield neighborhood-level insights. In \textbf{Activity 04}, we analyzed tweets from the police's Twitter account to gauge citizen feelings of insecurity, but the data did not support any understanding of insecurity. This failure produced a key RtD insight: passive digital listening was insufficient in this context. Instead, we needed a proactive, material intervention. This led to a second requirements-gathering step \textbf{(Activity 05)}, resulting in the requirement for \textit{a hybrid tool that supports neighborhood policing by integrating incident data with citizen feedback, while remaining operationally lightweight and compatible with existing stakeholder workflows}.
\end{sloppypar}

\subsubsection{Prototyping and Iteration} \label{sec:prototyping-iteration}

\begin{figure}[ht]
  \centering
  \includegraphics[width=1.0\linewidth]{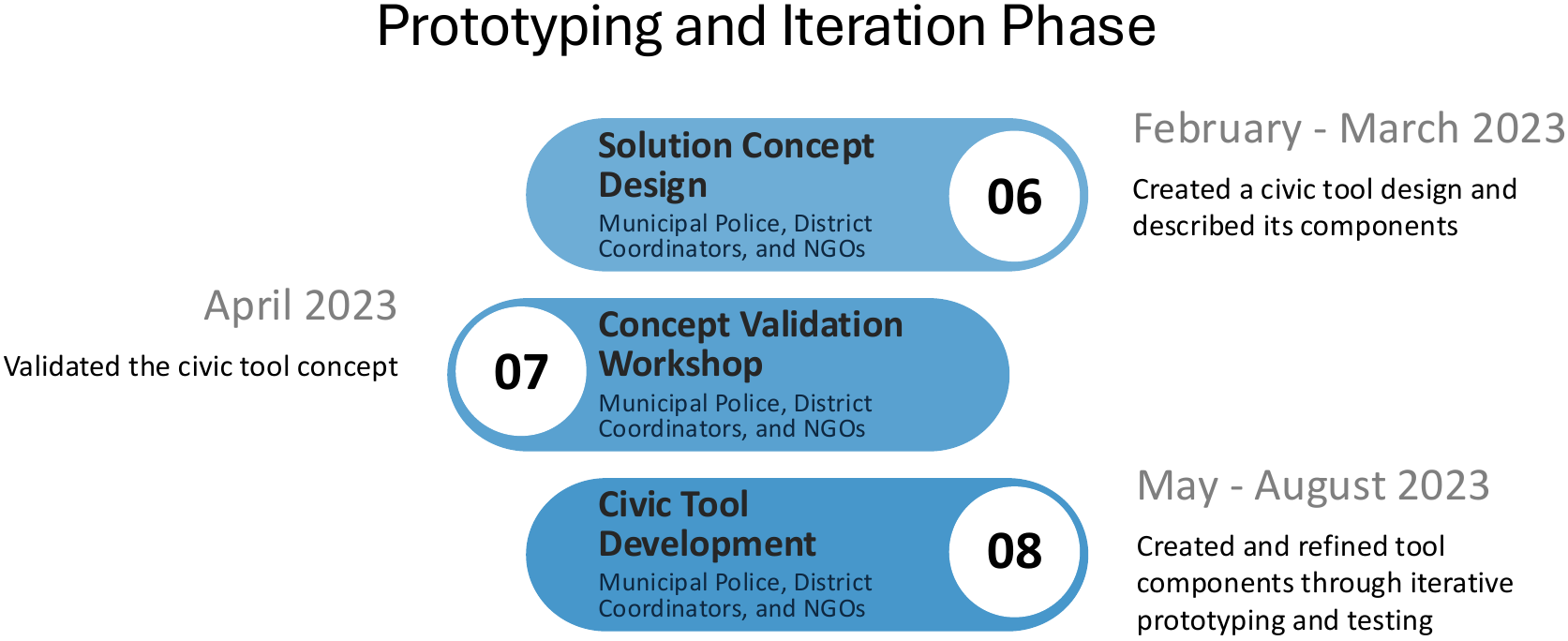}
  \caption{Activities of prototyping and iteration phase}
  \Description{Timeline diagram of the ``Prototyping and Iteration'' phase (February–August 2023) showing three activities: (06) Solution Concept Design with municipal police, district coordinators, and NGOs (February–March 2023); (07) Concept Validation Workshop with the same stakeholders (April 2023); and (08) Civic Tool Development with the same stakeholders (May–August 2023).}
  \label{fig:pd2}
\end{figure}

In the second phase (Figure~\ref{fig:pd2}), we formed a Community of Interest (COI) comprising a subset of participants from the design thinking workshop and organized a series of co-design activities, as it was difficult to convene all participants for every activity. The COI included police officers, NGO volunteers, a district coordinator (from the central district municipal office), and the municipal police IT specialist. Based on the results of previous activities, the research team developed a \textit{solution concept} \textbf{(Activity 06)} consisting of: (i) a three-part citizen survey administered in the field by police, municipal staff, and NGO volunteers; and (ii) visualization templates for police incident records and survey responses, presented via a web application. The concept targets neighborhood-level implementation within a district under the responsibility of the police Chief of District (CoD). Rather than relying on external researchers, the concept positioned the stakeholders as data collectors. This choice was not only logistical but also political: it leveraged existing relationships and trust within the local stakeholder network~\cite{Saldivar2019}.

After validating the concept \textbf{(Activity 07)} with the COI, we collaboratively prototyped each component \textbf{(Activity 08)}. During the prototyping of the survey mechanism, we negotiated the tensions of legitimacy: NGOs were assigned to ask about trust in police (to ensure honesty), while officers were assigned to ask about feelings of insecurity (to demonstrate care). The prototyping phase thus served as a negotiation of roles, defining who had the right to ask which questions.

\subsubsection{Validation and Demonstration} \label{sec:validation-implementation}

\begin{figure}[ht]
  \centering
  \includegraphics[width=1.0\linewidth]{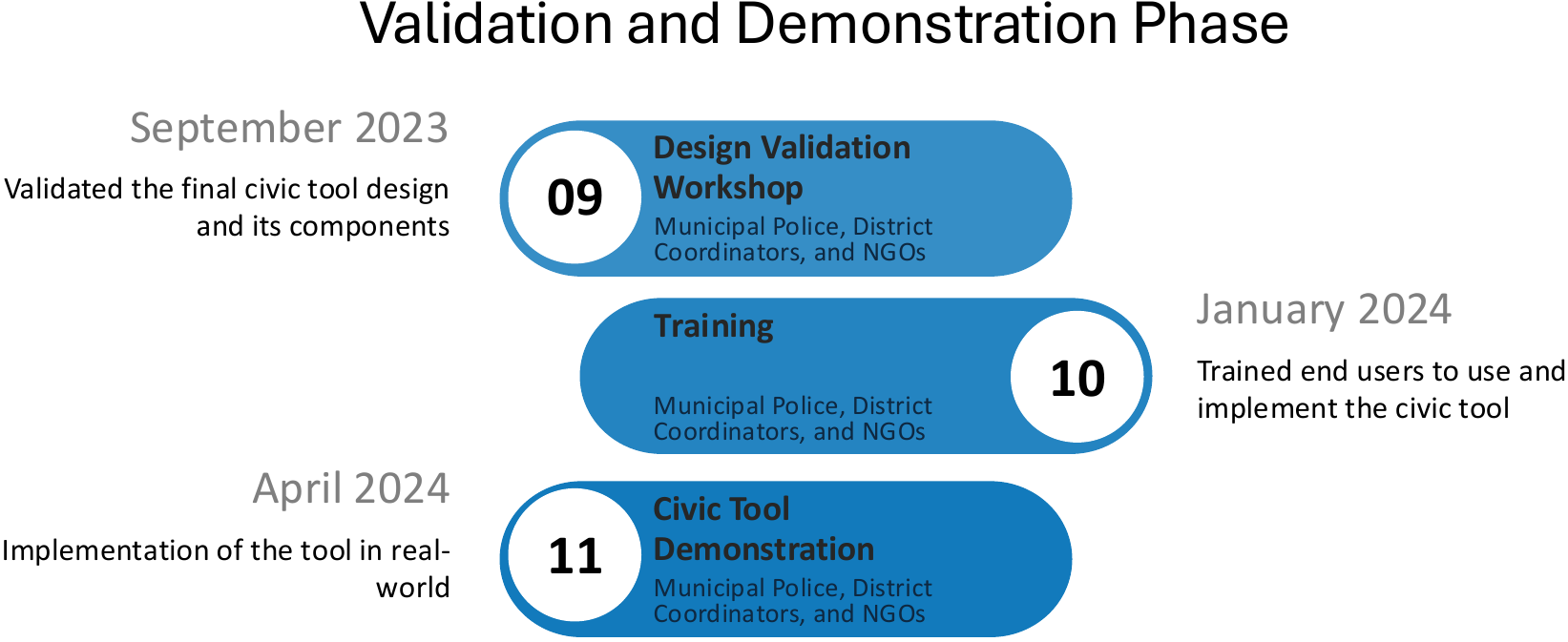}
  \caption{Activities of validation and demonstration phase}
  \Description{Timeline diagram of the ``Validation and Demonstration'' phase (September 2023–April 2024) showing three activities: (09) Design Validation Workshop with municipal police, district coordinators, and NGOs (September 2023); (10) Training with the same stakeholders (January 2024); and (11) Civic Tool Demonstration with the same stakeholders (April 2024).}
  \label{fig:pd3}
\end{figure}

The third phase (Figure~\ref{fig:pd3}) focused on the demonstration of the civic tool. After a \textit{Design Validation Workshop} \textbf{(Activity 09)} to test the finalized components and \textit{training sessions} \textbf{(Activity 10)}, we deployed the tool in one neighborhood in the city's central district \textbf{(Activity 11)}. The pilot involved six surveyors (two from each surveyor category) and one CoD.

During the deployment, the material and organizational frictions of the design became apparent. Patrol officers collected over 100 responses for their survey component, whereas the district coordinator and NGO volunteers collected only a small number. This divergence surfaced an RtD insight about the asymmetry between volunteer capacity and institutional capacity: even when a tool is designed to distribute participation, sustained engagement depends on time availability, role mandates, and visibility in public space. We concluded the demonstration with a joint stakeholder meeting to review results, share feedback, and assess the feasibility of adapting the tool to other districts.

\subsubsection{Assessment}\label{sec:methods-evaluation}

\begin{figure}[ht]
  \centering
  \includegraphics[width=1.0\linewidth]{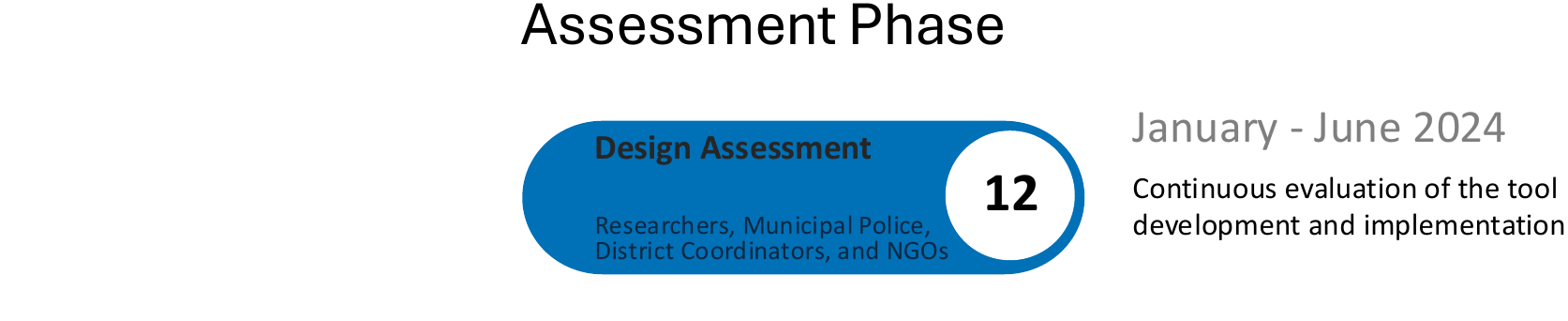}
  \caption{Activity of assessment phase}
  \Description{Timeline diagram of the ``Evaluation'' phase (January–June 2024) showing Activity 12 (``Design Assessment'') involving researchers, municipal police, district coordinators, and NGOs.}
  \label{fig:pd4}
\end{figure}

\begin{sloppypar}
We conducted a multi-method evaluation (\textbf{Activity 12}, Figure~\ref{fig:pd4}) consisting of (1) usability tests of the tool's components and (2) continuous assessment embedded into the process.
\end{sloppypar}

\paragraph{Tool Components}

We evaluated the final components through role-based usability sessions. Successful adoption was indicated by smooth task completion without interruptions: surveyors launched the field survey and recorded responses, while CoDs used the web application to analyze incident and survey data and save results. Participants worked independently and received minimal guidance only when needed. After the session, they completed a questionnaire. To assess the web application's perceived quality beyond task performance, we administered a short version of the Visual Aesthetics of Websites Inventory (VAWI)~\cite{moshagen2013} to measure design consistency, visual clarity, and attractiveness; the User Experience Questionnaire (UEQ)~\cite{martin2014} for attractiveness, efficiency, and satisfaction; and the Self-Assessment Manikin (SAM)~\cite{bradley1994} to capture emotional responses across pleasure, arousal, and dominance.

\paragraph{Process} \label{sec:methods-evaluation-development}

\begin{sloppypar}
Assessment is often omitted in civic technology and urban security projects because it is perceived as expensive, time-consuming, and too broad to be practical for practitioners~\cite{Saldivar2019, Crawford2023}. To address this, we integrated assessment into the tool's development and demonstration as a process of institutional learning and iterative refinement. Together with the municipal police, we used a logic model to define the tool's intended impact, outcomes, and mechanisms~\cite{McLaughlin1999}, and specified assessment indicators: ``stakeholder engagement'', ``citizen interaction'', ``levels of trust'', ``technological integration'', ``support of political leadership'', and ``tool sustainability''. For each indicator, we developed questions, defined data type (qualitative/quantitative), identified target groups, and set assessment timing (e.g., before, during, or after the demonstration). We applied these indicators across three phases—training, demonstration, and reception—to maximize data-collection opportunities while minimizing additional workload. We conducted semi-structured interviews with researchers, police, city officials (including Deputy Mayor), NGOs volunteers, and an independent representative from a Swedish city who were part of the process (Table~\ref{tab:participants-continious} in Appendix~\ref{app:participants}) and collected data across the three phases as follows: 
\end{sloppypar}

\begin{sloppypar}
\begin{itemize}
    \item \textbf{Training:} A post-training Likert-scale questionnaire assessed training effectiveness.
    \item \textbf{Demonstration:} We analyzed fieldwork diaries, media reports, survey data logs, and surveyor feedback to evaluate implementation.
    \item \textbf{Reception:} We reviewed interviews and a preliminary report on the tool's social acceptability to assess adoption and perception.
\end{itemize}
\end{sloppypar}

\section{Artifact: \textit{Par dro\v{s}u R\={\i}gu} civic tool} \label{sec:tool}

\begin{figure*}[ht]
  \centering
  \includegraphics[width=0.8\linewidth]{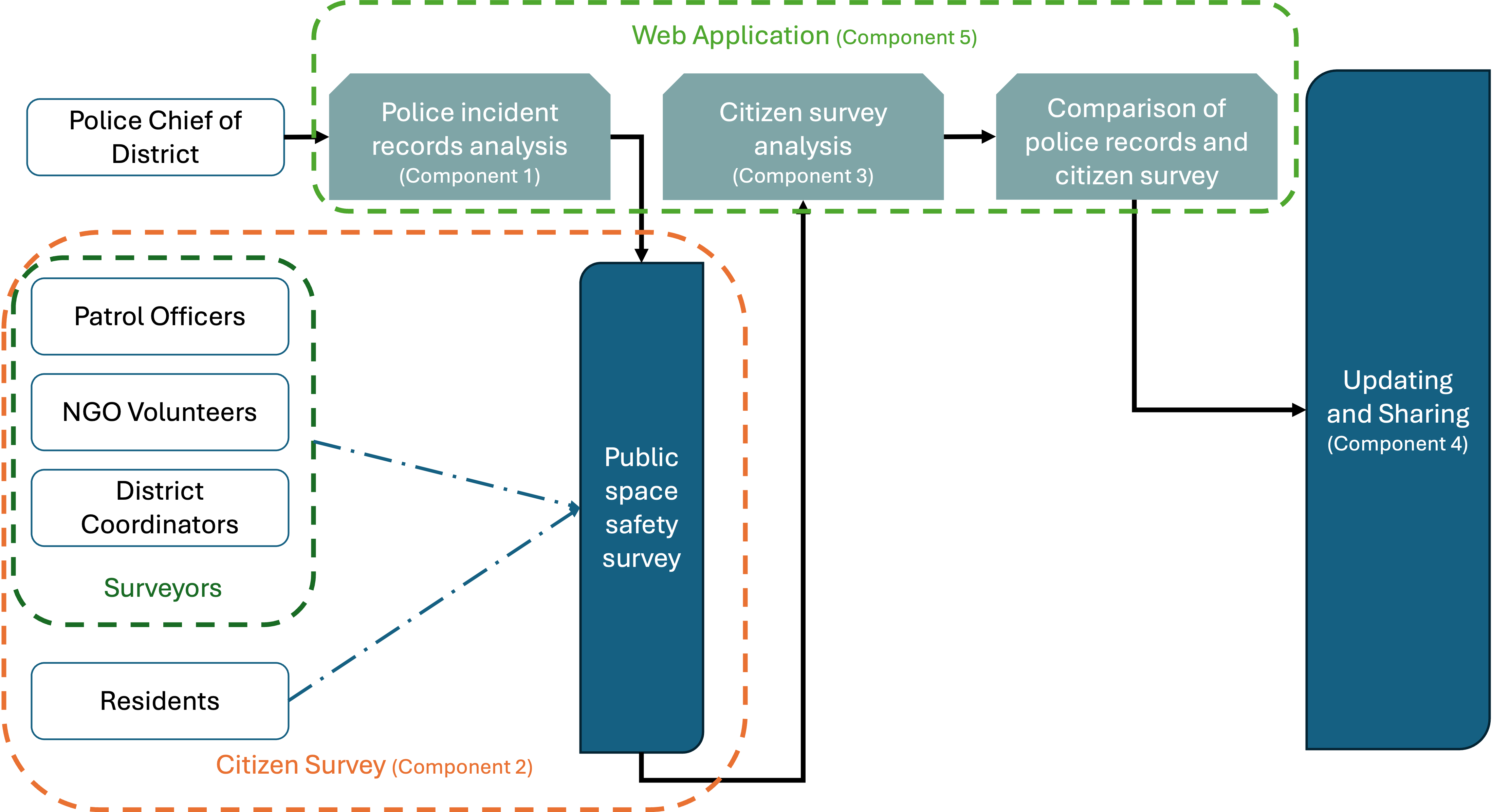}
  \caption{Components of the civic tool. The vertical span of the components represents the stakeholders (on the left-most pane) that are part of its interaction.}
  \Description{Block diagram showing stakeholders and components of the \textit{Par dro\v{s}u R\={\i}gu} civic tool. Left column lists stakeholders: Police Chief of District; surveyors (patrol officers, NGO volunteers, district coordinators); and residents. Middle and right show tool components and how they connect: Component 2 is a public-space safety survey conducted by surveyors with residents; Components 1 and 3 are police incident-record analysis and citizen survey analysis; a web application (Component 5) supports these analyses and comparison of police records with survey results; Component 4 is ``Updating and Sharing,'' which communicates findings and supports changes to policing practices. Dashed outlines group the survey activity and the web-application modules, indicating which stakeholders interact with each part.}
  \label{fig:riga-tool}
\end{figure*}

\textit{Par dro\v{s}u R\={\i}gu} civic tool is a socio-technical infrastructure that supports community-data- and interaction-based adaptation of neighborhood policing by bringing together officially recorded incidents and residents' lived experiences. The tool is intended to be run by the police Chief of District (CoD) twice per year (spring and autumn) as a four-stage cycle (Overview in Figure~\ref{fig:process-map} in Section~\ref{sec:intro}): \textit{Incident Analysis} (identifying trends and responses), \textit{Citizen Survey} (administered by patrol officers, the district coordinator, and NGO volunteers, followed by surveyor feedback), \textit{Survey Analysis} (interpreting citizen input in relation to incident records), and \textit{Updating Practices} (translating insights into changes and communicating outcomes). We translated this cycle through five components (Figure~\ref{fig:riga-tool}): (i) incident-record analysis, (ii) field surveying, (iii) survey analysis, (iv) updating and sharing, and (v) a web application that supports coordination and data handling across stages. User manuals and training guidelines accompany each component. Across these components, the tool supports (i) situated interactions of surveyors with residents in public space, (ii) human-data interaction as CoDs interpret incident and survey data through the web application, and (iii) institutional interactions between stakeholders. We positioned the tool as a boundary object that supports coordination across stakeholder groups~\cite{Lee2007} and as infrastructure for situated reflection and iterative learning~\cite{Peer2023, Crivellaro2019}, rather than as a system that automates decisions.

\subsection{Police Incident Records Analysis}

\begin{figure}[ht]
  \centering
  \begin{subfigure}{0.48\textwidth}
    \includegraphics[width=\linewidth]{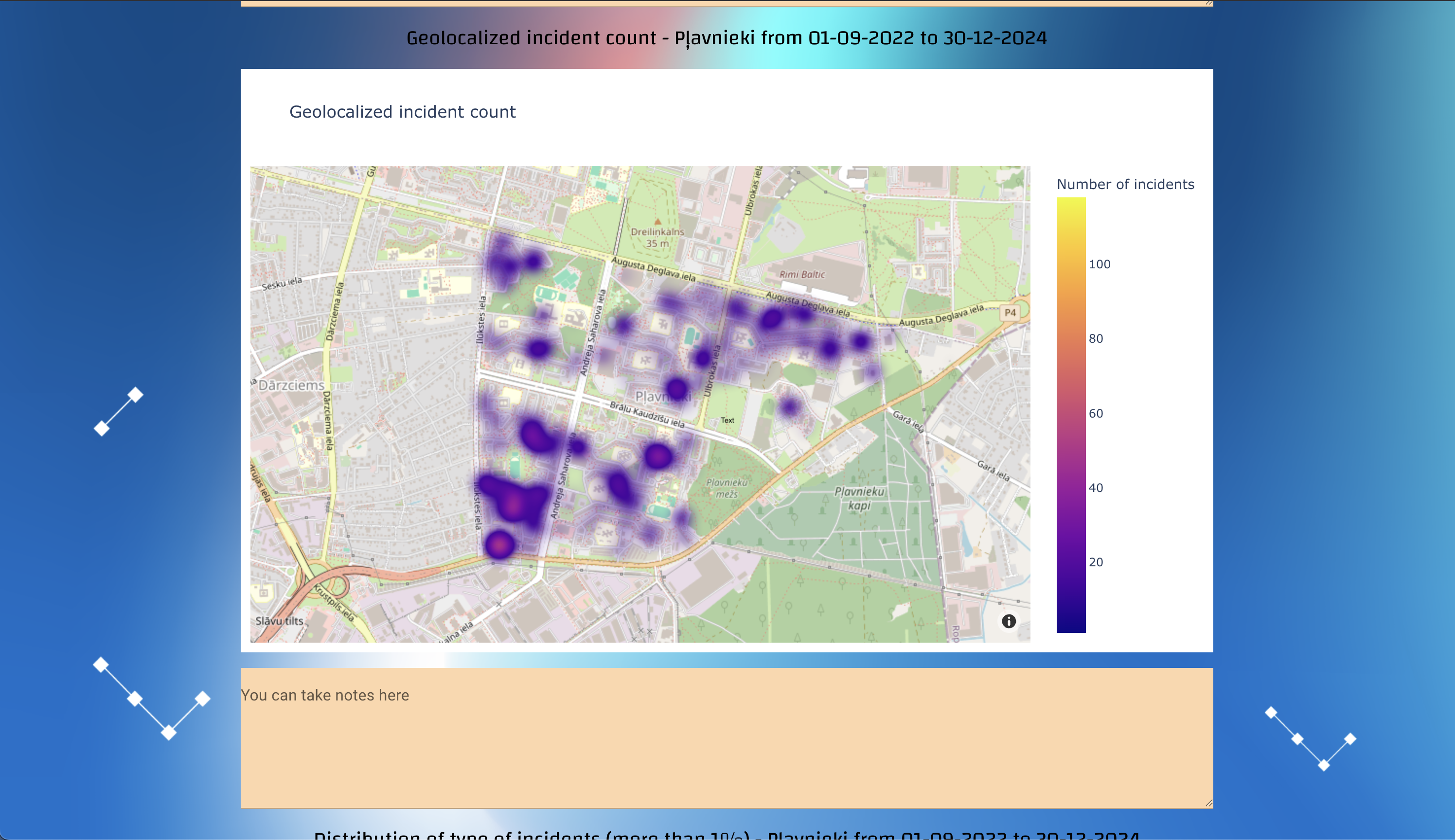}
    \caption{Report for incident record analysis - Geolocalized incident count.}
    \Description{Screenshot of the page of the web application dedicated to the incident record analysis. It shows the Geolocalized incident count at each address as a heat map imposed on the city map.}
    \label{fig:webapp-incident-analysis-2}
  \end{subfigure}
  \hfill
  \begin{subfigure}{0.48\textwidth}
    \includegraphics[width=\linewidth]{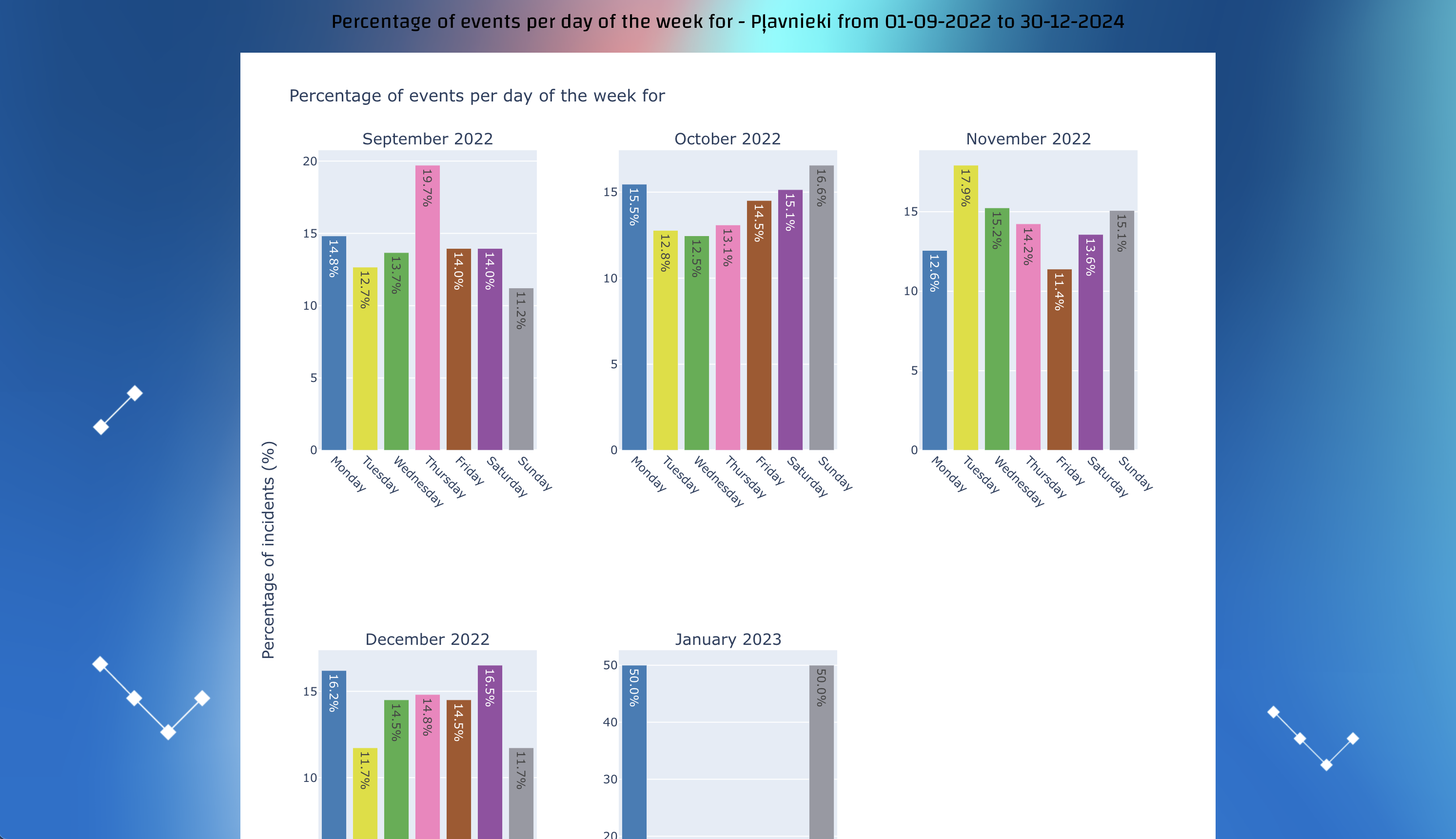}
    \caption{Report for incident record analysis - Percentage of events per day of week.}
    \Description{Screenshot of the page of the web application dedicated to the incident record analysis. It shows the barplot with the proportion of incidents reported per day of the week across each Month.}
    \label{fig:webapp-incident-analysis-3}
  \end{subfigure}
  
  \caption{Sample Screenshots of Incident Records Analysis (Created using synthetic data)}
  \label{fig:web-app-indicent-analysis}
\end{figure}

The first component responds to a persistent gap identified during early engagements with the municipal police: while data on reported incidents were systematically collected, it remained underutilized for informing tactical decisions at the neighborhood level. Incident records include fields such as date, time, location, incident type, reporting source, and follow-up actions. CoDs upload anonymized exports from internal police systems, and the civic tool generates ten interactive visualizations (Figure~\ref{fig:web-app-indicent-analysis}), including temporal distributions (hour/day/month), spatial heat maps, and response metrics. These visualizations are not prescriptive; they represent an institutional view of safety and help CoDs contextualize citizen feedback.

\subsection{Citizen Survey}

Early in the co-design process (Activity 06), we determined that a purely digital survey (e.g., a web link) would fail to reach marginalized residents or build trust. Instead, we co-designed the second component as a role-based neighborhood survey mechanism. The survey captures three dimensions of public safety---feelings of insecurity, perceived risks, and community cohesion---and, to balance visibility and access across social groups, each dimension is administered by a different group (patrol officers, district coordinators, or NGO volunteers; Table~\ref{tab:surveys}). We differentiated surveyor roles to navigate local legitimacy and power dynamics (e.g., reluctance to criticize the police directly) and to leverage each group's access to different resident segments. This is a deliberate interaction design intervention: it repurposes the patrol encounter, typically associated with enforcement or fines, into an interaction of listening and care. The NGO-administered part can support candid responses about trust, and district coordinators capture the views of residents already engaged with civic services. The survey questions (Appendix~\ref{sec:citizen-surveys}) were adapted from validated instruments in prior research on public safety and feelings of insecurity~\cite{Emily2010Feelings, Reid2020Perceptionsmodel}.

\raggedbottom
\begin{table}[ht]
  \caption{Division of surveys between surveyors}
  \label{tab:surveys}
  \Description{Table mapping each surveyor group to where they recruit respondents and the survey topic they administer. Patrol officers survey citizens passing through the streets about feelings of insecurity. District coordinators survey visitors at the district municipal center about resident cohesion and neighborhood disorder. NGO volunteers survey participants at NGO activities and public events about anxiety of crime, trust in police, and perceived risk of harm.}
  \begin{tabular}{p{0.2\linewidth}p{0.33\linewidth}p{0.33\linewidth}}
    \toprule
    \textbf{Surveyor} & \textbf{Respondents} & \textbf{Topic} \\ \midrule
    Patrol\newline officers & Citizens passing\newline through the streets & Feelings of insecurity \\
    
    District\newline coordinators & Visitors of the district municipal center & Cohesion among the\newline residents and the\newline disorder in the\newline neighborhood \\
    
    NGO\newline volunteers & Participants in NGO\newline activities and public\newline events & Anxiety of crime,\newline trust in police, and\newline perceived risk\newline of harm \\
  \bottomrule
\end{tabular}
\end{table}

\begin{sloppypar}
Each survey part was designed to take under ten minutes and to be administered in person through online forms managed by the police. Direct interaction allows for clarification of questions and cultivates rapport, leading to more accurate and candid responses~\cite{Saldivar2019, Johnson2021}. In-person surveys also ensure better control over data quality, reducing the risk of fraudulent or duplicate entries and supporting stratified sampling for a representative understanding of the population~\cite{Golsteijn2016}. This approach combines the benefits of face-to-face interaction with the efficiency of digital data collection. In addition, the criteria to select the surveyors were defined during co-design workshops. Patrol officers are selected based on their communication skills, familiarity with Crime Prevention Through Environmental Design (CPTED) principles~\cite{Cozens2015}, and prior community-facing experience, while maintaining gender balance. District coordinators are delegated by the district municipal offices. NGO volunteers are selected based on their effective communication skills and familiarity with the neighborhood. At the end of the survey, surveyors fill an online questionnaire (see Appendix~\ref{sec:surveyors-feedback}) to reflect on their experiences and suggest improvements.
\end{sloppypar}

\subsection{Citizen Survey Analysis}

\begin{figure}[ht]
  \centering
  \begin{subfigure}{0.48\textwidth}
    \includegraphics[width=\linewidth]{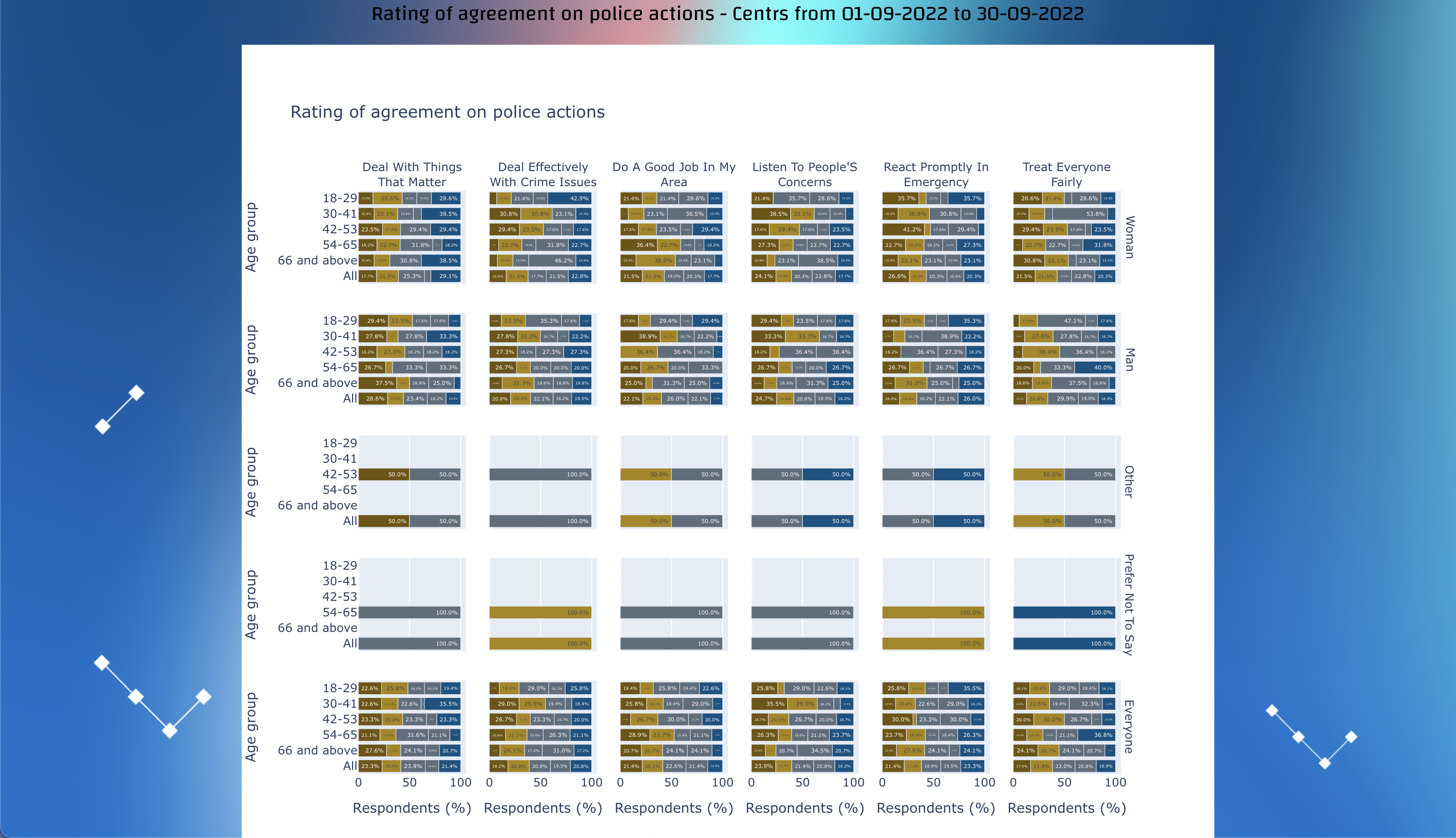}
    \caption{Report for citizen survey analysis - Rating of agreement on police actions.}
    \Description{Screenshot of the page of the web application dedicated to the incident record analysis. It shows the rating of respondents' trust in police actions.}
    \label{fig:webapp-citizen-survey-analysis-1}
  \end{subfigure}
  \hfill
  \begin{subfigure}{0.48\textwidth}
    \includegraphics[width=\linewidth]{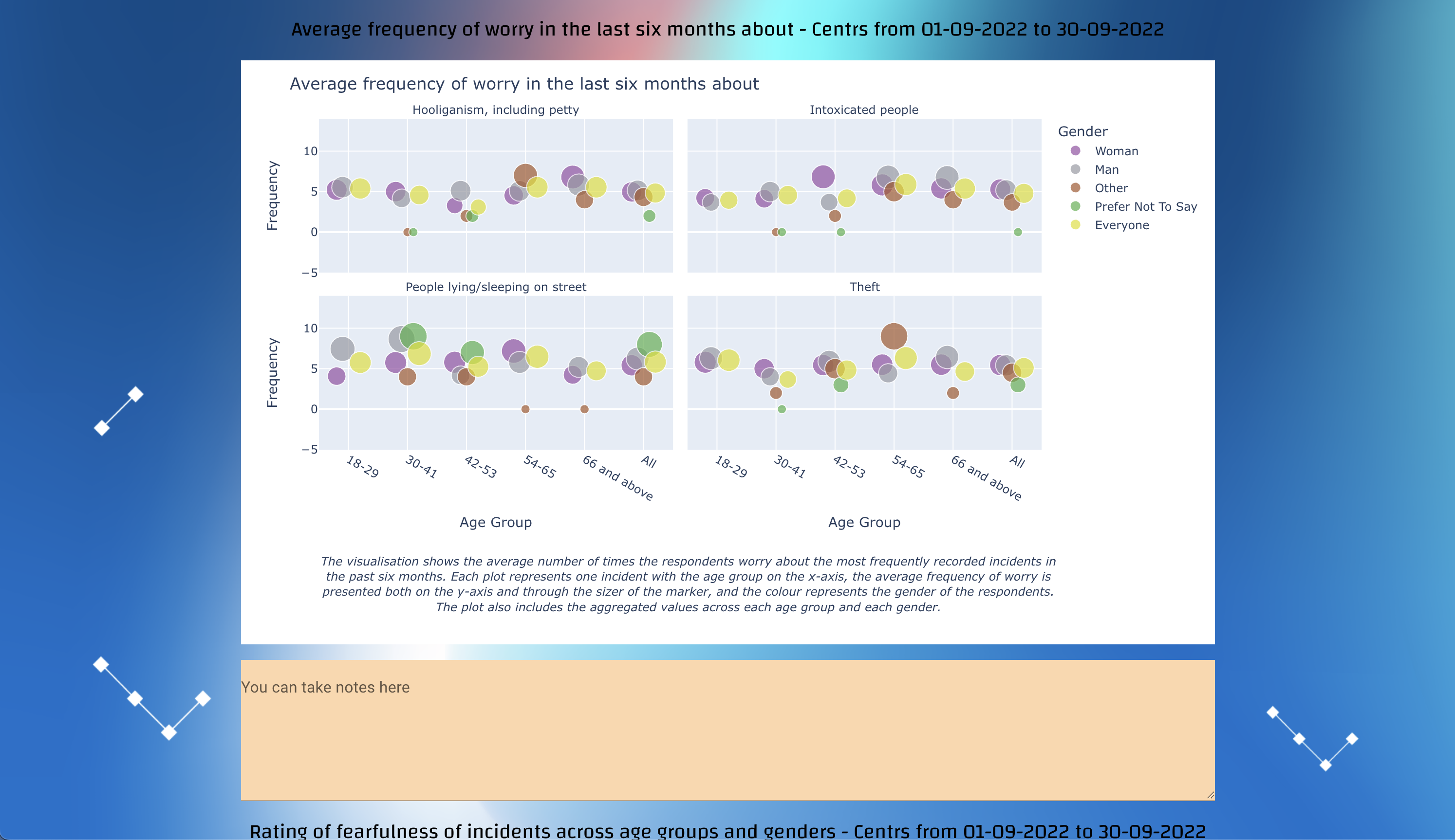}
    \caption{Report for citizen survey analysis - Average frequency of worry in the last six months.}
    \Description{Screenshot of the page of the web application dedicated to the incident record analysis. It shows the average number of times respondents worried about frequently reported incidents in the last six months.}
    \label{fig:webapp-citizen-survey-analysis-2}
  \end{subfigure}
  
  \caption{Sample Screenshots of Citizen Survey Analysis (Created using synthetic data)}
  \label{fig:webapp-citizen-survey-analysis}
\end{figure}

\begin{sloppypar}
The third component helps CoDs understand citizens' subjective experiences with security, trust, and cohesion---dimensions not captured by incident data alone. Figure~\ref{fig:webapp-citizen-survey-analysis} shows sample visualizations of the survey data on the web application. The aggregated responses of three survey dimensions are visualized. The feelings of insecurity survey allows police to identify specific public areas and streets perceived as unsafe, even in the absence of incident reports. It also shows the fear of using public transport and preferred patrol locations. A second cluster of charts focuses on neighborhood social dynamics, based on the community cohesion survey (e.g., trust among neighbors and collective action in response to suspicious activity). The third survey, measuring perceptions of police trustworthiness, responsiveness, and perceived fairness, is also visualized and disaggregated by age and gender. These visualizations represent the lived experience view and help officers identify community groups with lower trust levels and initiate targeted outreach. The web interface allows CoDs to view incident and survey data side-by-side. This juxtaposition transforms the dashboard into a boundary object~\cite{Dalsgaard2014}: it makes discrepancies visible. For example, a CoD might see that a specific park has zero recorded crimes (Safe in Data) but a high avoidance score (Unsafe in Sentiment). These tensions of differences in data are mediated through the dashboard's commentary box feature, where CoDs can annotate the data with their interpretations.
\end{sloppypar}

\subsection{Updates and Sharing}

The civic tool was not designed to automate decision-making or enforce a fixed data-action pipeline. Instead, it supports a flexible process of interpretation, negotiation, and action, enabling CoDs to triangulate between citizen perceptions, incident reports, and operational priorities to update their community policing steps. These adjustments include redistributing patrols to areas where perceptions of insecurity or disorder have increased, reprioritizing incident responses based on citizen concerns, and scheduling community meetings in neighborhoods with low cohesion or trust scores. CoDs also share their analyses beyond the district. They report to the strategic planning office of the Riga municipal police when data suggests city-wide trends; if issues fall outside police purview, the office forwards feedback to the relevant municipal departments. In parallel, CoDs meet regularly with district coordinators and NGO partners to present selected analyses, commentary, and planned actions, along with surveyor feedback, and to jointly discuss implementation adjustments.

\begin{figure}[ht]
  \centering
  \begin{subfigure}{0.48\textwidth}
    \includegraphics[width=\linewidth]{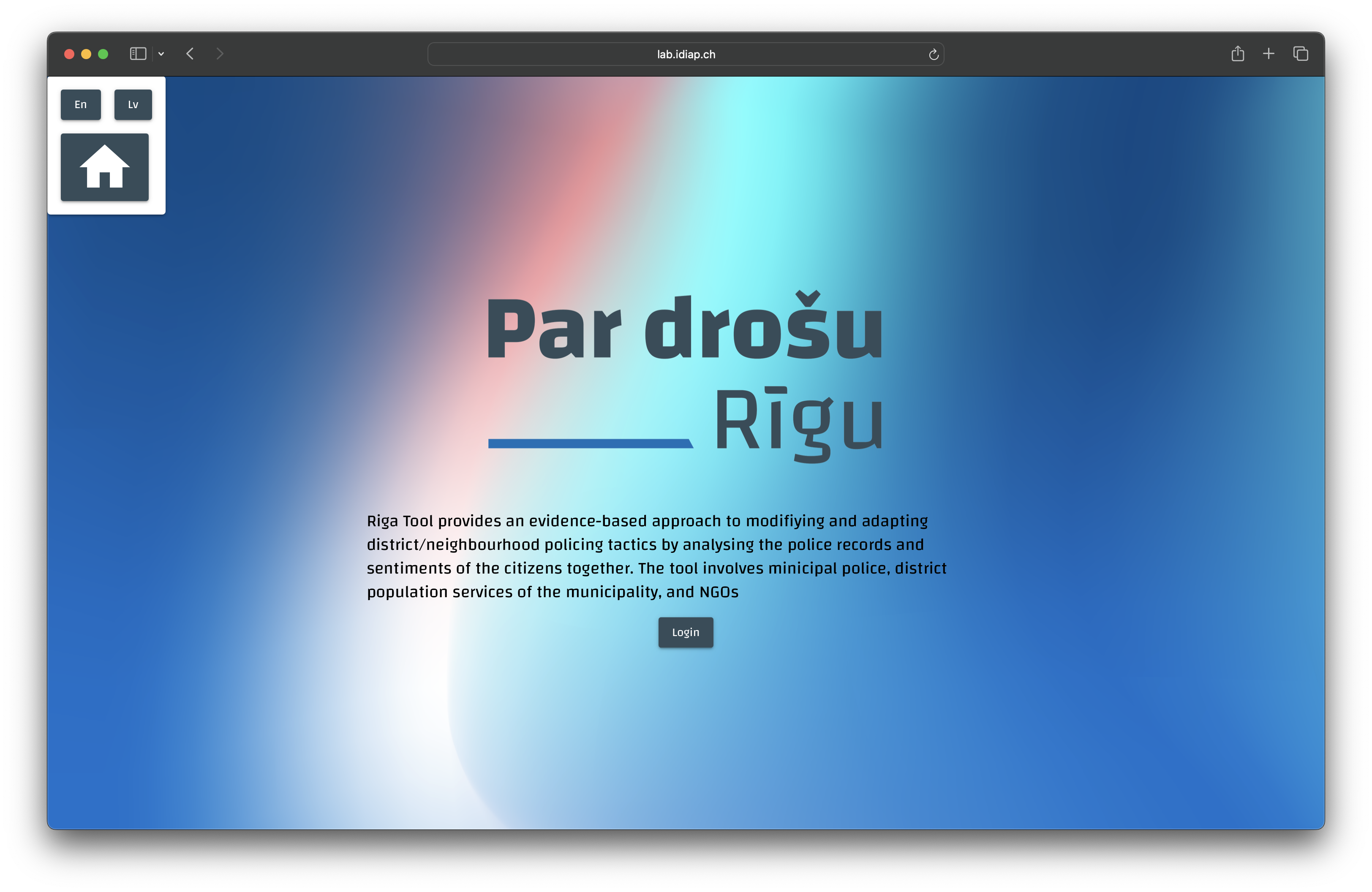}
    \caption{Landing page.}
    \Description{Screenshot of the landing page of the web application. It includes the name of the tool and a short description on a purple and blue background.}
    \label{fig:webapp-landing-page}
  \end{subfigure}
  \hfill
  \begin{subfigure}{0.48\textwidth}
    \includegraphics[width=\linewidth]{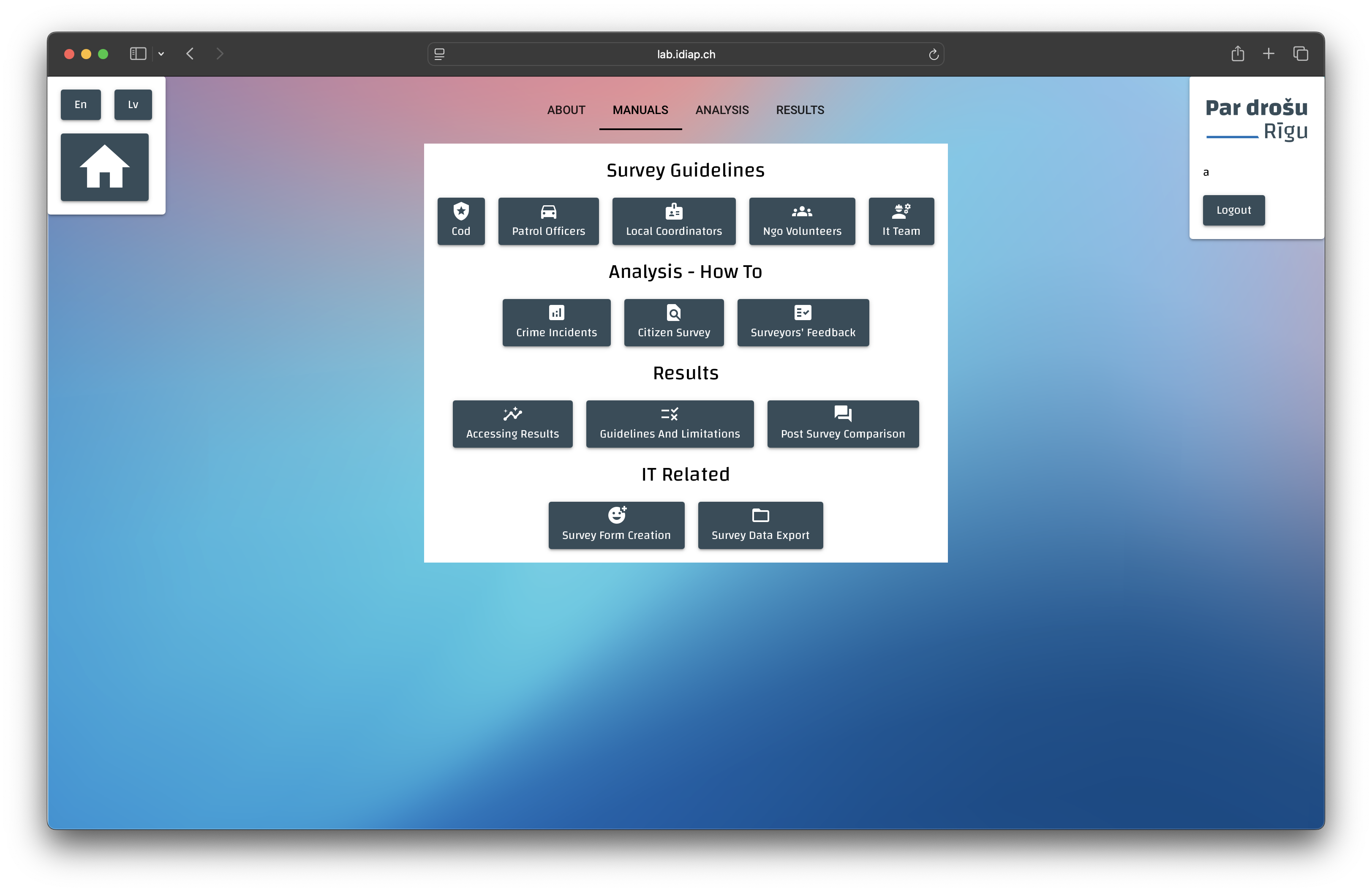}
    \caption{Page for manuals of the civic tool.}
    \Description{Screenshot of the page of the web application dedicated to the manuals of the civic tool. It shows several manuals to help implement the civic tool.}
    \label{fig:webapp-incident-analysis}
  \end{subfigure}
  \hfill
  \begin{subfigure}{0.48\textwidth}
    \includegraphics[width=\linewidth]{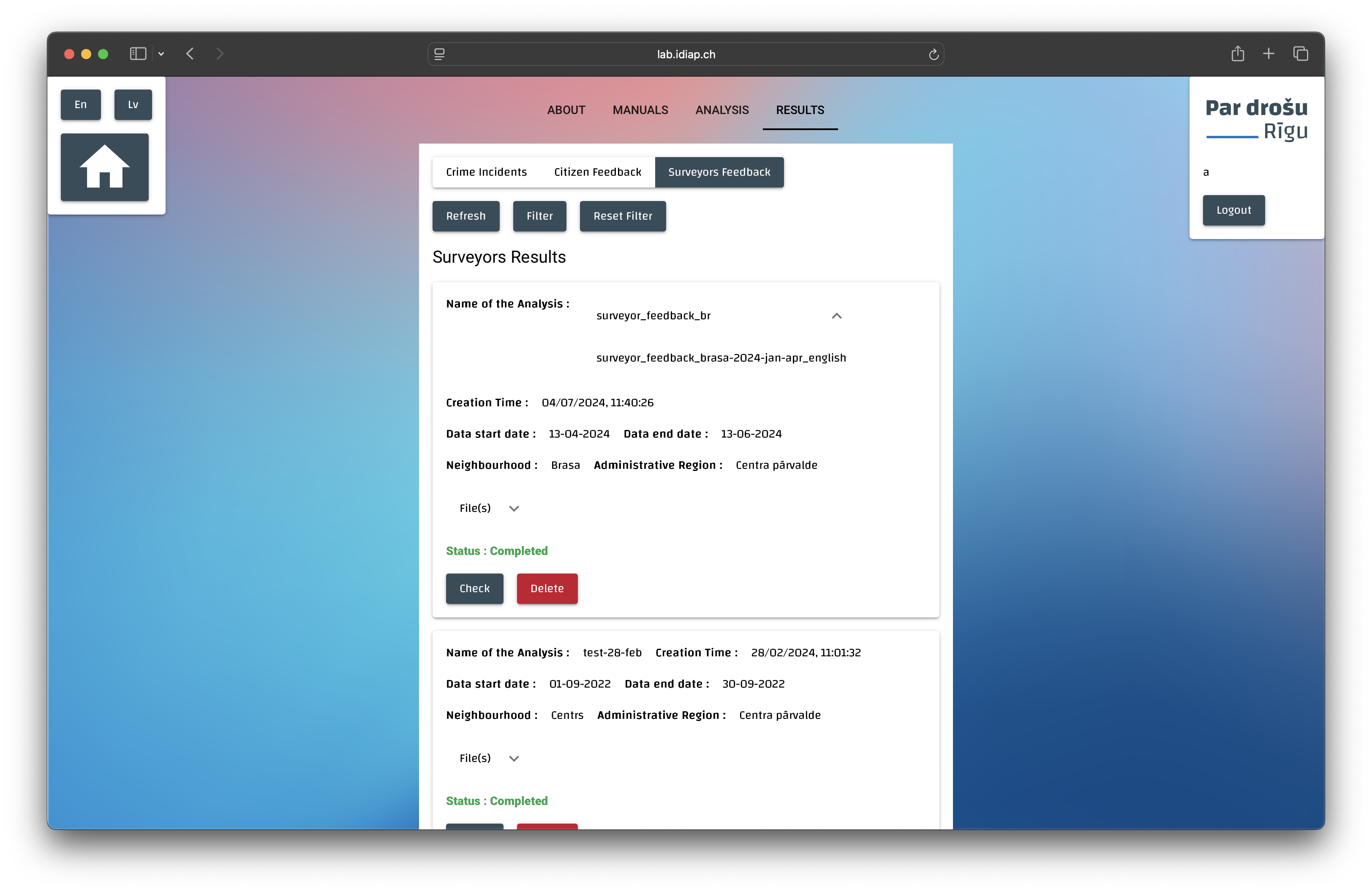}
    \caption{Page for the results of surveyors' feedback.}
    \Description{Screenshot of the page of the web application dedicated to the results of the surveyors' feedback. It shows two different files with relevant information, such as dates, neighborhoods, and titles, which can be clicked and opened.}
    \label{fig:webapp-surveyors}
  \end{subfigure}
  
  \caption{Interfaces of the Web Application}
  \label{fig:web-app}
\end{figure}

\subsection{Web Application}

The web application is the interface through which CoDs coordinate, analyze, and archive incident records, survey responses, and surveyor feedback across the cycle (Figure~\ref{fig:web-app}). It is designed to be lightweight and usable by CoDs without reliance on external technical staff, which was essential given the limited IT capacity. Each visualization in the web app allows CoDs to annotate findings, and those qualitative entries are archived alongside graphs, which can be exported as reports for dissemination.

\section{Design Assessment} \label{sec:evaluation}

\subsection{Tool Components}

\subsubsection{Citizen Survey}

Four participants (patrol officers and district coordinators; Table~\ref{tab:cit-survey} in Appendix~\ref{app:participants}) participated in the usability testing of the surveying component. All participants reported that interaction with the online survey form was straightforward (Figure~\ref{fig:surveyor-tool-eval} in Appendix~\ref{app:toolcomp-evaluation}). However, they anticipated challenges related to field conditions, including limited responsiveness from residents (participants estimated collecting 20--100 responses over two weeks), adverse weather, and difficulty reaching all age groups. Overall, participants viewed the surveying component as usable and feasible, while noting practical constraints that could affect coverage.

\subsubsection{Web Application}

\begin{sloppypar}
Five participants (Table~\ref{tab:web-app} in Appendix~\ref{app:participants}) tested the web application. We first assessed perceived visual design and professionalism (layout, color, and overall coherence). Participants generally described the interface as clean and professional (Figure~\ref{fig:vis-as} and \ref{fig:cod-tool-eval} (in Appendix~\ref{app:toolcomp-evaluation})). They noted that the layout and visual elements contributed to a coherent and aesthetically pleasing experience. One participant perceived a mismatch with their expectations, suggesting further refinement may be needed as needs evolve. Secondly, we assessed interaction experience across ease of use, efficiency, and clarity (Figure~\ref{fig:ux}). Most participants rated the experience positively, particularly for ease and perceived supportiveness, both of which matter for integration without extensive training.
\end{sloppypar}

\begin{figure}[ht]
  \centering
  \begin{subfigure}{0.48\textwidth}
      \includegraphics[width=1\linewidth]{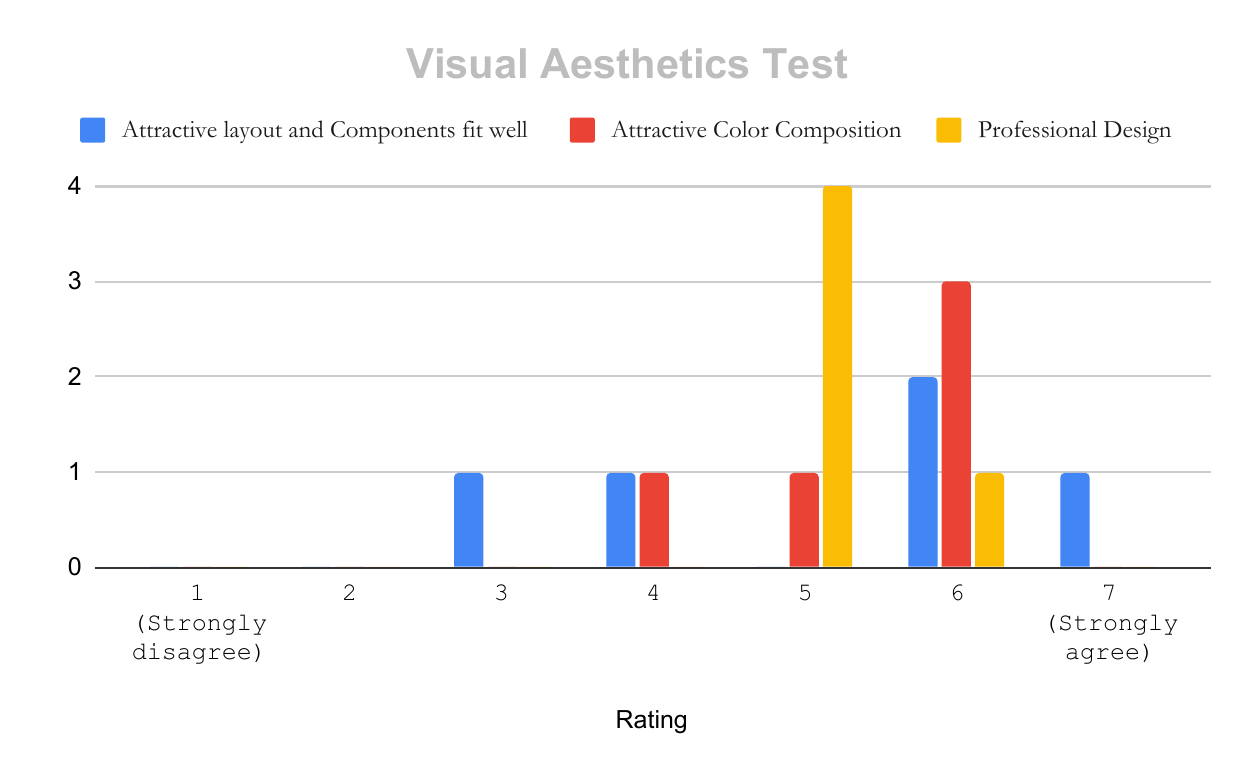}
      \subcaption{Perceptions of visual aesthetics.}
      \Description{Bar chart showing participant feedback on visual design elements of the web application.}
      \label{fig:vis-as}
  \end{subfigure}
  \hfill
    \begin{subfigure}{0.48\textwidth}
      \includegraphics[width=1\linewidth]{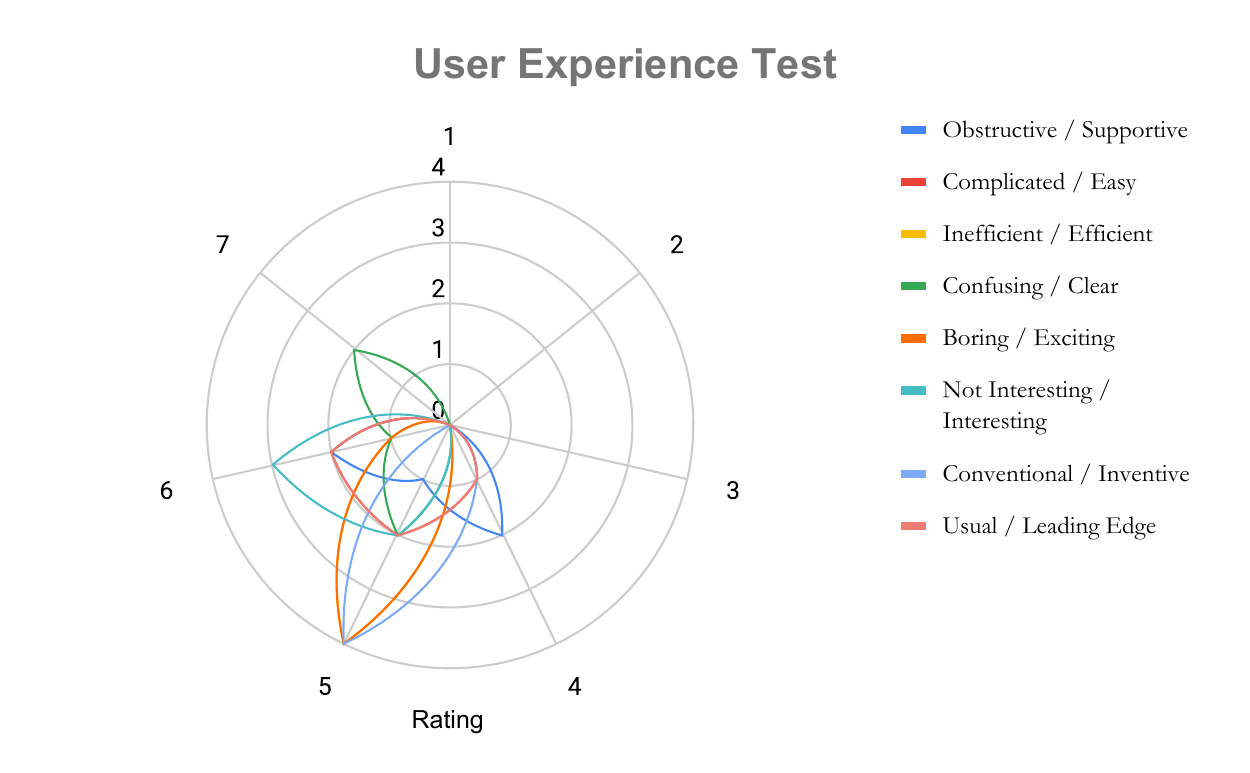}
      \subcaption{User experience ratings for the web application.}
      \Description{Radial chart showing user feedback on the overall experience with the web application.}
      \label{fig:ux}
  \end{subfigure}

  \caption{Web application design evaluation with 5 users.}
  \label{fig:feedback-user-tests}
\end{figure}

\subsubsection{Integration with existing workflows}

Participants judged integration on two aspects: \emph{data utility} and \emph{workflow fit}. CoDs valued the survey data and dashboard analytics, but reported limited interaction with surveyors---signaling uneven readiness across roles. Participant \textit{WA2} noted that streamlined data gathering could significantly enhance their efficiency, while \textit{WA5} highlighted visualizations as prompts for reallocating patrols and cameras. Most testers agreed the tool could surface new, neighborhood-level insights even when individual features felt familiar. \textit{WA4} indicated that by using this new system, they will be able to carry out broader information gathering and save time. Older participants scored aesthetics and ease slightly lower than their younger colleagues, but age trends were modest, and no gender effects emerged, given the small sample size. Some participants used the web application on mobile and faced technical issues during testing, as it was not primarily designed for mobile screens.

Participants assessed integration along two dimensions: data utility and workflow fit. CoDs valued the survey data and dashboard analytics, but reported limited interaction with surveyors, suggesting uneven readiness across roles. Participant \textit{WA2} noted that streamlined data gathering could improve efficiency, while \textit{WA5} described the visualizations as prompts for reallocating patrols and cameras. Most participants agreed the tool could surface neighborhood-level insights even when individual features felt familiar. \textit{WA4} expected the system to support broader information gathering while saving time. Older participants rated aesthetics and ease slightly lower than younger colleagues; given the small sample, these differences should be treated as indicative rather than conclusive. Some participants used the web application on mobile devices and encountered issues, consistent with the application not being designed primarily for small screens.
 
\subsection{Process}\label{sec:eval-continuous-assessment}
 
\subsubsection{Stakeholder Engagement}

\begin{sloppypar}
The pre-existing relationships between the stakeholders---CoDs meeting quarterly with local NGOs and ongoing coordination between district coordinators and police---were reported by interviewees to have facilitated stakeholder engagement. In particular, to obtain willingness to participate in the process from the earliest stage, the multi-stakeholder collaboration was described as engaging. \textit{CA1} noted: ``different perspectives, different opinions,'' and \textit{CA2} added: ``very fruitful, personally.'' Patrol officers' willingness to participate and engage with citizens was repeatedly perceived as an important success. Their participation was recognized as having positive implications on the city's capacity to act as a stakeholder within the wider European environment. The deputy mayor (\textit{CA4}) noted that it gives Riga and other European cities an early chance to share challenges and build resilience together. However, the interviews suggested that formal co-ownership of the tool would have enabled greater levels of synergy in terms of responsibility (who can do what, when, and how). For example, NGOs could have conducted the survey during their activities to reduce the burden. 
\end{sloppypar}

\subsubsection{Citizen Interaction}

Citizen interaction was described by all interviewees as a positive outcome. Interviewees reported that the civic tool created new opportunities for communication with residents. \textit{CA1} said: ``The patrol officers got different perceptions [on] how the people living in the area, people on the street feel, understand, and see the territory..., they understood... There are a lot of people like normal people, not only violators, perpetrators...'' This development was interpreted as a positive signal to improve community-police relations more broadly. This increased citizen interaction was desired but not replicated across all stakeholders. District coordinators and NGOs experienced practical challenges in obtaining survey responses. \textit{CA5} said: ``I work in an office and thought ... visitors will come... But they didn't, so the last day I also went on the streets and got some responses.'' \textit{CA7} cautioned: ``obtaining few responses in a very quick time to overcome limited interaction can skew the data.'' The importance of citizen interaction is therefore important for the data quality obtained through the survey. Patrol officers' results show this is attainable with the civic tool, while the other surveyors' findings underscore the importance of stakeholder engagement.

\subsubsection{Levels of Trust}

Based on the positive levels of citizen interaction reported by patrol officers, interviewees believed that trust between citizens and police had improved in their perception. \textit{CA2} described a shift toward ``human conversation between officers and inhabitants.'' \textit{CA1} noted that even if residents declined to respond, the civic tool will be meaningful: ``... like win-win because you go to the people, you talk with them and it strengthens everything... destigmatize the police profession, because from... some kind of street bureaucrat or like person who is writing you the penalty, you move to the category of people who care, who want to listen and who want to speak.'' The deputy mayor (\textit{CA4}) said the civic tool could be an example of how to strengthen trust between public authorities and citizens: ``If you look at this crisis in lack of confidence in public authorities in Europe as a whole, but also in Latvia, within that distrust, law enforcement agencies are those institutions which have most of the trust actually... so we have to use this [success] to demonstrate that the public sector is genuinely interested in what our people think and how they feel.''

\subsubsection{Technology Integration}

Interviewees separated two technical parts of the tool: the citizen survey and the web application. The survey also succeeded as a data-collection effort, as noted by the CoD (\textit{CA3}): ``the goals were achieved, new and relevant data on the security situation in the neighborhood were obtained.'' Moreover, \textit{CA7} noted that it ``was innovative for the city as it provides new and cost-effective collection of neighborhood-specific data on feelings of insecurity along with feedback from police officers on the ground, police records, and a visual assistant for decision-making.'' However, the web application had diverging reactions. The police expected an automated pipeline to merge existing policing data and new survey data and provide interpretations. Other interviewees said: ``when we developed the tool, we all thought that it would be a help for decision'' (\textit{CA6}) and ``since the conceptualization, we always use the word assist policing decision-making... never framed as predictive tool'' (\textit{CA7}). The language of the different stakeholders (both in terms of native language and data literacy language) proved to have created a disjuncture between the different partners. Nevertheless, the police (\textit{CA1}) noted that the ``perspectives the respondents ... are totally different from the statistics... we got totally different results from results we get daily [from the police incident analysis].'' This demonstrated that while the web application ``didn't fully meet the expectations'' of the municipal police (\textit{CA7}), it enhanced existing working practices and provided greater capacity. Despite the gap, external observers found the approach transferable; a Swedish city representative (\textit{CA8}) expressed interest in implementing a similar tool in their city because of the tool's ability to increase levels of communication and trust.

\subsubsection{Political Leadership}

The support of political leadership was important for enabling sufficient resources to be allocated to the tool's implementation and adoption. The deputy mayor's interest in ``developing the police force in general'' (\textit{CA2}) was deemed to contribute to stakeholder buy-in and engagement, providing the political backdrop against which changes could be justified and supported through the tool. Simultaneously, the perceived success of the tool in increasing interaction between police and citizens facilitated further political support. As the deputy mayor (\textit{CA4}) noted: ``[residents] complain about ... feelings of insecurity... It doesn't help if we say your complaints are not valid because we don't have data, or it's a very secure area... it will be interpreted as arrogance and ... make the situation of distrust even worse. Although we invest city resources in the civic tool... in the end, I think it pays back in a more integrated and resilient society.''

\subsubsection{Tool Sustainability}

\begin{sloppypar}
Interviewees framed sustainability as part of a longer-term ambition to improve trust and engagement between the police (as representatives of public authority) and residents. They viewed the tool’s lightweight components and low-cost, survey-based engagement as enabling continued use beyond the initial pilot. In terms of resourcing and institutionalization, the deputy mayor (\textit{CA4}) noted: ``We will see how to multiply this experience also to other areas... develop this learning culture within the organization, readiness and openness to learn new things...'' The tool was also described as a mechanism for influencing working practices and organizational culture within the police, by illustrating potential benefits for ``procedural justice, trust, legitimacy and communication'' and beginning to ``change the mindset'' and ``orthodox'' culture that can discount the value of citizen interaction (\textit{CA1}).
\end{sloppypar}

Following the initial pilot, the police continued to use and adapt the tool in another neighborhood. Over three weeks, two patrol officers conducted the survey ($N{=}100$) at two public locations (a bus stop near a local supermarket and a dog-walking area). At both locations, officers reported collecting information that was not documented in the police incident records. Based on survey responses, a supplementary action to respond to perceptions of insecurity on nighttime public transit was incorporated for the second location. The findings were subsequently shared with community members and local officials at the Nightlife conference in 2025.


\section{Discussion} \label{sec:discussion}

\subsection{Co-Design for Public Safety}

At the outset, we jointly framed the problem as a data-alignment issue: official incident statistics and responses from the city's biennial public safety survey often diverged, creating a gap between recorded data and citizens' perceptions. Police treated safety primarily as a function of response time and efficient deployment. Citizen data was therefore approached through a service-optimization lens, with limited attention to relational dimensions of policing such as legitimacy, trust, and perceived fairness~\cite{bradford2010trust}.

\begin{sloppypar}
The design thinking workshop (Activity 02) surfaced that safety was experienced differently across stakeholder groups: some emphasized enforcement and patrolling, others inclusion, and others physical road infrastructure. This led to vague outcomes and required additional co-designing activities, which is difficult in time-constrained or resource-limited environments~\cite{Kirsi2011, Cruciani2023}. This tension illustrates a core trade-off: co-design is necessary to align a civic tool with diverse needs and expectations, but it also demands sustained commitment and flexibility. In our case, this process pushed the project beyond incident mapping toward representing perceptions and social relationships. Concretely, we moved toward a structured yet participatory data collection approach through a situated three-part survey. This shift aligns with work on publics and participation, where technologies help constitute publics and make issues actionable~\cite{Dantec2012}, and offers an alternative to control- or surveillance-oriented approaches to public safety~\cite{Park2017}. The \textit{Par dro\v{s}u R\={\i}gu} civic tool was designed with the assumption that citizens may be willing to engage and can contribute local knowledge to public safety efforts~\cite{Erete2015, Dickinson2021}. Based on this experience, we recommend early introductory meetings and workshops to establish a shared understanding of terms, goals, and tool concepts across perspectives for future civic tool development projects~\cite{Sistrunk2024}.
\end{sloppypar}

\begin{figure*}[ht]
\centering
\includegraphics[width=\linewidth]{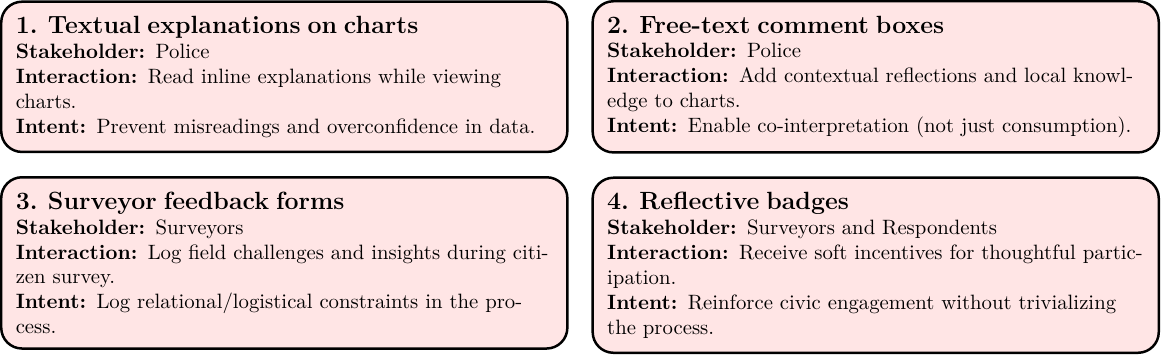}
\caption{Interaction mechanisms addressing digital literacy and civic feedback.}
\label{fig:interaction-mechanisms}
\Description{Four labeled boxes describing interaction mechanisms in the civic tool: (1) textual explanations on charts for police to reduce misreadings; (2) free-text comment boxes for police to add contextual reflections; (3) surveyor feedback forms for surveyors to log field challenges and insights during the citizen survey; and (4) reflective badges for surveyors and respondents as soft incentives for participation.}

\end{figure*}

\subsection{Interaction Design for Trust} 

We do not pretend to solve the public safety challenges through the co-designed civic tool. Instead, it provides an infrastructural alternative with a recurring interaction process for reflection in which interpretations and actions can be discussed and iteratively revised~\cite{Zhang2020}. Several design intentions remained aspirational and were not fully realized in practice. As prior work notes~\cite{Ming2021}, civic tools can suffer from underuse, nonresponsiveness, and one-way communication, which limit their potential. Our work underlined the complexities of interaction design research in the policing context and shows the criticality of balancing the time investment with practical outcomes for building similar civic tools. Incremental approaches that prioritize small-scale, iterative changes may be more suitable than large, one-time interventions. They allow stakeholders to gradually build trust, adapt to institutional constraints, and accommodate their evolving relationship~\cite{Schaap2021, Corbett2018, Corbett2021}. 

\begin{sloppypar}
Our co-design process was marked not only by collaborative intentions but also by persistent asymmetries in power, resources, and legitimacy. The municipal police participated consistently across activities, provided access to institutional data, and framed the initial problem. Their role as institutional lead, availability, and clear operational stakes gave them greater influence over functional priorities and intended uses. In contrast, NGOs and district coordinators participated more episodically, constrained by time, resources, and limited institutional obligation, consistent with prior findings \cite{Merkel2004}. To stabilize participation, we established a Community of Interest (COI; Phase 2/Section~\ref{sec:prototyping-iteration}), which helped formalize roles and sustain feedback channels. Disparities in digital literacy and familiarity with design research and local language also shaped how feedback was articulated and interpreted. Recognizing that technical features can reinforce stakeholder power dynamics~\cite{Manuel2020, Crivellaro2019}, we implemented interaction mechanisms explicitly oriented toward accountability and interpretability. Figure~\ref{fig:interaction-mechanisms} summarizes the resulting design features, including plain-language explanations and comment boxes to support interpretation and capture implementation feedback, as well as badges acknowledging participation.
\end{sloppypar}

\subsection{Local Factors Influencing Civic Tool Demonstration}

\begin{sloppypar}
Our assessment suggests that stakeholders responded positively to the tool's usability and to increased opportunities for community-police engagement. Patrol officers, due to their routine street presence, institutional legitimacy, and potential power asymmetries in police-citizen interactions~\cite{bradford2010trust, Skogan2006}, collected substantially more survey responses. In contrast, NGO volunteers and district coordinators, who we expected to have strong community proximity, collected few responses, largely due to limited time and lower visibility or authority in the public space. This highlights an institutional friction in relying on volunteering for civic infrastructures. While the design intent was to democratize data collection, in practice, the police were the only stakeholder group with sufficient capacity to scale this form of interaction. This constraint underscores the importance of aligning civic tools with existing workflows. While such asymmetry raises concerns about response bias~\cite{Ming2021}, we did not identify scalable alternatives within the constraints of the setting. More broadly, this episode surfaces the persistence of unequal participation in co-design and civic technology projects~\cite{Crivellaro2019, Skogan2006}. We therefore intended prototypes and artifacts as boundary objects that are shared resources that remain interpretable across stakeholder groups while supporting coordination and alignment~\cite{Lee2007, Dalsgaard2014}. Despite these mechanisms, complete symmetry remained elusive, consistent with prior accounts~\cite{Holone2013}.
\end{sloppypar}

Another barrier concerned data interpretation. While CoDs could interpret individual charts during prototyping, they struggled to integrate multiple charts during the demonstration, a difficulty also reported in prior work~\cite{Edwards2010, Haque2024}. This ties into broader issues of information, digital, and data literacy. These are sets of skills and capabilities to discover and understand information, make sense of and analyze data~\cite{Shields_2005}, and utilize digital tools~\cite{Tinmaz2022}. Within the RtD inquiry, the pilot surfaced additional design insights: although CoDs had prototyped the web application, hands-on use generated new requests (e.g., automated interpretation), indicating that material engagement can function as a probe that elicits latent needs. These emergent needs illustrate the value of RtD for working with non-designers and public institutions, where use becomes a site of shared discovery and informs next co-design iterations~\cite{Zimmerman2014, Gaver2012, Edwards2010}.

\subsection{Implications for RtD with Public Institutions (RQ2)}

Our study suggests four implications for RtD when designing civic tools with public institutions. First, civic tools must be designed to function as boundary objects~\cite{Lee2007} where they should be understandable and valuable to multiple stakeholders, even when each group interprets them through a different lens. Second, we should anticipate and accommodate ``meaningful inefficiencies''~\cite{Gordon2019}. Public institutions often prioritize reliability, accountability, and risk mitigation over rapid innovation. Slower decision cycles and literacy differences can appear as friction, but they can also become conditions for building trust and legitimacy, which are prerequisites for sustained engagement~\cite{Saldivar2019, Corbett2018, Corbett2021}. Third, civic tools must be designed as infrastructuring for compatibility with existing data formats, networks, and everyday work practices~\cite{Crivellaro2019, Saldivar2019, Peer2023}. In our case, this meant using police record exports in their original form, without additional work, leveraging existing stakeholder networks for surveying, and making mediation visible through plain-language explanations, annotations, and participation incentives (Figure~\ref{fig:interaction-mechanisms}). Making this mediation obvious smoothed friction and integration, aligning with accounts of infrastructuring and human work that make infrastructures work~\cite{Monteiro2013, Cruciani2023, Ludwig2018, Dickinson2021}. Fourth, RtD in public institutions should plan for organizational buy-in and capacity-building, not only the artifact. Even a well-designed tool can fail if it clashes with power structures or if primary users are not empowered to operate it~\cite{Dantec2013, Skogan2006}. In our case, repeated co-design engagements and training helped establish initial practices for using the civic tool. 

\begin{sloppypar}
Future initiatives should similarly budget time and resources for cultivating relationships, aligning expectations, and developing institutional capability to sustain the tool. Our case shows that, in policing, RtD is a balance between broadening participation and designing forms of negotiation within hierarchical structures and varied perceptions of legitimacy. These insights extend prior RtD accounts by showing how hierarchy, legitimacy, and interpretive responsibility become important design concerns in community-engaged research with police. This distinction was also reflected in how police compared RtD to typical software procurement. Riga municipal police contrasted the RtD/co-design process with prior collaborations with software vendors (e.g., to develop the Electronic Event Log to log incident data and a mobile application for residents' reporting). They described these earlier collaborations as a ``unidirectional process'' in which the police specified a solution and vendors proposed implementation and pricing. In contrast, the RtD work required shared problem framing, engagement with prior cases and evidence, and negotiation with NGOs as intermediaries for inhabitants’ perspectives, illustrating how ``meaningful inefficiencies'' can function as capacity-building rather than overhead.
\end{sloppypar}

\subsection{Limitations and Future Work}

\begin{sloppypar}
Our study has limitations that suggest directions for future work. First, the findings are bound to the specific socio-political context of Riga~\cite{Cruciani2023} and are not easily generalizable. Integrating civic tools into decision-making remains contingent on long-term institutional willingness to act on citizen input~\cite{Manuel2020, Crivellaro2019}. Second, due to local access constraints, citizen involvement in co-design was mediated through NGOs~\cite{Holone2013} rather than direct participation. Future work should explore the inclusion of additional stakeholders, such as local businesses, schools, and social organizations, to address gaps in data literacy, local knowledge, and power~\cite{Johnson2018}. Third, juxtaposing community-generated data with official records creates interpretative challenges. While we view this friction as productive and a negotiation space, the civic tool leaves the interpretation to the police; future iterations could incorporate collaborative annotation to support joint interpretation across stakeholders. Lastly, a long-term evaluation could examine more directly how the civic tool operates as a boundary object and increases trust in practice.
\end{sloppypar}

\section{Conclusion} \label{sec:conclusion}

\begin{sloppypar}
In this paper, we detail a Research through Design (RtD) inquiry in which we co-designed a civic tool to support community-data-integrated policing in the Baltic city of Riga, with limited community-police collaboration. The \textit{Par dro\v{s}u R\={\i}gu} civic tool functioned both as an RtD outcome and as a probe that surfaced infrastructural tensions in integrating citizen feedback into policing practices. Across the paper, we contribute an empirical RtD account, the \textit{Par dro\v{s}u R\={\i}gu} civic tool artifact, and reflective insights from mixed-methods evaluation. Our findings inform RtD accounts of public-sector design by showing how a civic tool can mediate between institutional data practices and residents' lived experiences.
\end{sloppypar}

Addressing \textbf{RQ1 (How RtD facilitates negotiation between multiple stakeholders)}, we described a two-year co-design process and the resulting civic tool that brings together police records and citizen feedback through a situated citizen survey. In the implemented configuration, CoDs (police chiefs) engage with incident and survey data through a web application, while stakeholders engage residents through in-person surveys. Across the process, we observed recurring frictions around trust-building, data literacy, and stakeholder alignment. These findings underscore the need for flexible probes, iterative refinement, and organizational arrangements that allow a civic tool to remain responsive to local constraints and community needs.

\begin{sloppypar}
Addressing \textbf{RQ2 (What interaction qualities support trust and data materialization)}, our assessment surfaced different orientations toward evidence: one based on operational data and another on lived experience. The final design represents a negotiated hybrid that juxtaposes police records with citizen perceptions and supports interpretation through lightweight interaction mechanisms. Although the co-design process was time- and coordination-intensive, it was important for the legitimacy of the interaction in this context. Stakeholders reported that the tool supported shifts in officer-resident interactions from enforcement toward listening and care, and political leaders framed it as a step toward rebuilding confidence in public institutions. Overall, our results suggest that the civic tool operated not only as a decision-support artifact but also as a boundary object and infrastructural arrangement that enabled ongoing multi-stakeholder collaboration and trust-building.
\end{sloppypar}

\begin{acks}
This work was supported by the IcARUS project, funded by the European Union's Horizon 2020 research and innovation programme under grant agreement No. 882749. We would like to thank all the participants in the city of Riga and our partners at the European Forum for Urban Security (EFUS) for their support. We would like to give a special thanks to Raimonds Nitišs (Riga Municipal Police) for his invaluable help in the organization of the project and coordination of logistical aspects.
\end{acks}

\bibliographystyle{ACM-Reference-Format}
\bibliography{references}


\appendix
\section{Co-Design Process Details} \label{app:pd-process}

We follow Zimmerman et al.'s~\cite{Zimmerman2007} RtD conceptualization, reflecting on how the design process unfolded; what socio-technical configurations emerged; how they addressed local needs; and how the resulting practices might transfer to other contexts. The activities are summarized in Table~\ref{tab:phases}. 

\begin{table}[ht]
  \centering
  \caption{Summary of activities and outputs}
  \label{tab:phases}
  \Description{Table summarizing four phases of the project, listing activity ranges and their main outputs: Activities 01--05 (Framing, alignment, requirements) produced understanding of context, needs, and stakeholders; Activities 06--08 (Prototyping and iteration) produced the civic tool; Activities 09--11 (Validation and demonstration) involved real-world implementation; Activity 12 (Design assessment) examined the process, design, and demonstration of the civic tool.}
  \begin{tabular}{p{0.46\linewidth}p{0.4\linewidth}}
    \toprule
    \textbf{Activity} & \textbf{Output} \\ \midrule
    01--05 / Framing, Alignment, and Requirements & Understood the context,\newline needs and stakeholders \\
    
    06--08 / Prototyping and\newline Iteration & Developed \textit{Par dro\v{s}u R\={\i}gu} civic tool (Section \ref{sec:tool}) \\
    
    09--11 / Validation and\newline Demonstration & Real-world\newline implementation of \textit{Par\newline dro\v{s}u R\={\i}gu} civic tool \\

    12 / Design Assessment & Examined the process,\newline design, and demonstration of \textit{Par dro\v{s}u R\={\i}gu} civic tool (Section \ref{sec:evaluation}) \\
  \bottomrule
\end{tabular}
\end{table}

\subsection{Framing the Problem, Aligning Stakeholders, and Gathering the Requirements}

\begin{figure}[ht]
  \centering
  \includegraphics[width=1.0\linewidth]{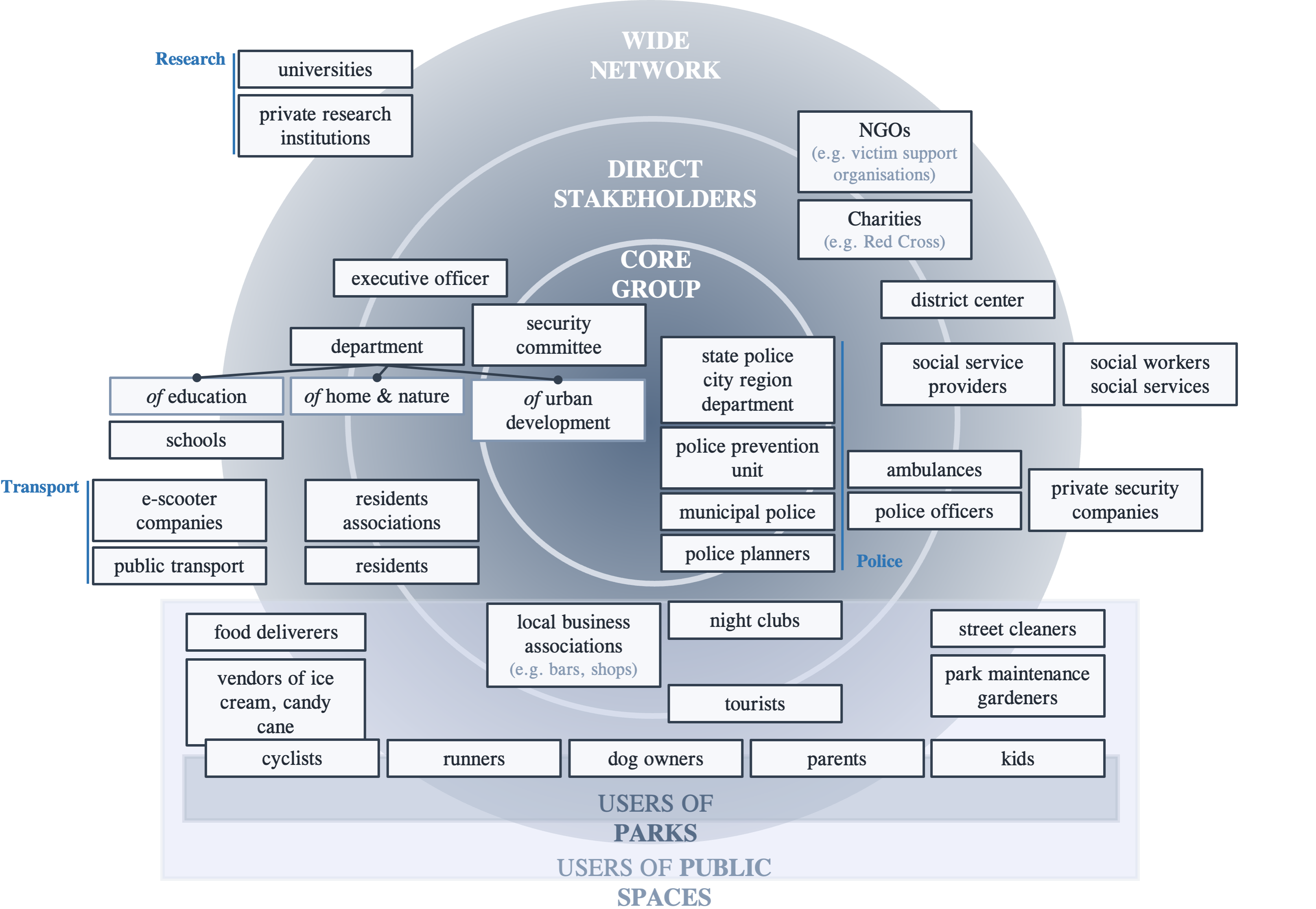}
  \caption{Stakeholder mapping, which resulted from the preparatory workshop}
  \Description{Diagram with all stakeholders directly or indirectly involved with them of safe public spaces in the Riga. This includes three levels: a core group, direct stakeholders, and a wide network. The core group includes law enforcement agencies and a security committee. Direct stakeholders include local businesses, NGOs, users of parks, and public spaces. The wide network encompasses public transport, researchers, and social workers.}
  \label{fig:stakeholders}
\end{figure}

The preparatory workshop (activity 01) with seven participants (Table~\ref{tab:participants-preparatory} in Appendix~\ref{app:participants}) discussed the needs and opportunities for Riga. After discussing the state-of-the-art urban security practices for maintaining safe public spaces, the Riga municipal police highlighted existing issues in their policing practices. They expressed the existing mechanisms in the city to aid the police, including a mobile application and regular meetings with local NGOs, present several gaps that do not allow for an effective adaptation of police practices based on citizens' needs. They noted discrepancies between a biennial survey of citizens and crime statistics, revealing inconclusive perceptions of neighborhood safety and public space risks, and received complaints about safety concerns beyond their control, such as inadequate street lighting in parks.
The preparatory workshop concluded with a stakeholder mapping to identify potential stakeholders for the co-design process, as shown in Figure~\ref{fig:stakeholders}.

Then, 18 participants from 4 stakeholder groups took part in the design thinking workshop (activity 02) that took place in June 2022. The workshop was organized by a facilitator trained in design thinking methodology, and conducted in Latvian, and its results were later transcribed into English. The challenges identified were diverse and did not provide a clear direction for solutions.
This made it necessary to refer back to the municipal police to better understand their needs and challenges. We regrouped with three representatives of municipal police in September 2022 to address gaps left by the design thinking workshop through a questionnaire (activity 03). This led to a tweet analysis (activity 04) of the police's Twitter account. Most discussions, however, centered on traffic and parking issues---unrelated to the identified challenge---so this aspect was not included in the civic tool design. The municipal police noted that, while hearing complaints about traffic/parking issues on Twitter was important, they needed diverse information for a better assessment of the ground situation and to help them update their practices. So, we regrouped for a requirements-gathering phase with the municipal police (activity 05). 

\subsection{Prototyping and Iteration}

After defining requirements, the next phase focused on iteratively developing and refining the civic tool. This involved translating the abstract design goals into functional components through prototyping sessions and feedback loops with the Community of Interest (COI). They provided regular input to the co-design process and the future users of the civic tool. Between February and March 2023, the core elements of the tool---its data sources, interface features, and use cases---were proposed as a concept (activity 06). To validate this concept, we organized a concept validation workshop (activity 07) in April 2023 with COI. The workshop was guided by structured feedback forms that captured participants' impressions of the system's usefulness, clarity, and feasibility. Participants were generally positive, citing the tool's potential to bridge current information gaps and support community engagement efforts. 

Based on this feedback, we developed and prototyped the tool components between May and August 2023 (activity 08). During this period, each component was iteratively tested and refined in collaboration with stakeholders. 
In each iteration, stakeholders were invited to test components relevant to their role---for example, the police focused on dashboard usability, while NGO members reviewed survey language and engagement strategies. This targeted feedback approach ensured that revisions were grounded in real-world use conditions and stakeholder expectations. 

\subsection{Validation and Demonstration}

\begin{figure}[ht]
 \centering
  \includegraphics[width=1.0\linewidth]{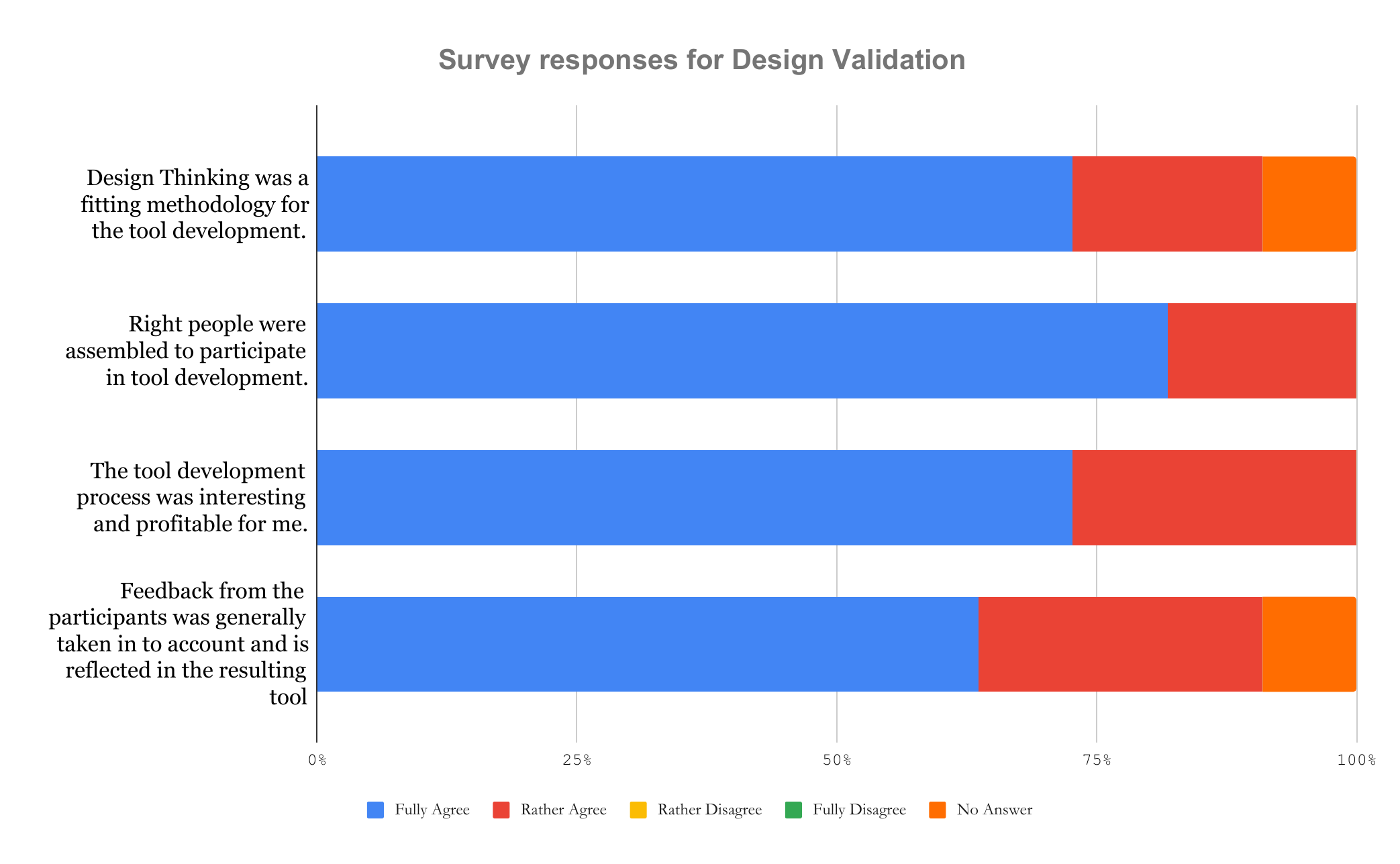}
  \caption{Survey responses for Design Validation}
 \Description{Horizontal bar chart showing the results of a questionnaire on the three following aspects: The Right people were assembled to participate in tool development; The tool development process was interesting and profitable for me; Feedback from the participants was generally taken into account and is reflected in the resulting tool. The results of all three questions indicated a prevalence of ``fully agree'' responses.}
 \label{fig:design-vald-survey}
\end{figure}

The completed tool components were tested at a design validation workshop in September 2023 (activity 09) with eleven members of the COI. COI also reflected on the co-design process itself. We collected feedback through anonymous surveys, which revealed a strong sense of participant ownership and satisfaction with the development process. As shown in Figure~\ref{fig:design-vald-survey}, most participants either fully or partially agreed that the right stakeholders had been involved, that their feedback had been taken into account, and that the process had been engaging and productive. Some participants emphasized the importance of inter-institutional collaboration and continuous dialogue with citizens to ensure long-term success.

After the validation, the users (already part of COI) were trained on using the tool in January 2024 (activity 10). The training covered technical aspects of installing and operating the web application, as well as guidelines for interpreting visualizations and planning follow-up actions. Surveyors---including patrol officers, NGO representatives, and district coordinators---were trained in administering the three-part citizen survey. The training emphasized ethical engagement, data quality, and role-specific survey delivery techniques.

The first demonstration of the tool took place in April 2024 in a selected neighborhood (activity 11). During the first two weeks, surveyors collected responses from citizens using online survey forms. Each group of surveyor focused on different dimensions of safety and trust. In the third week, CoDs conducted an analysis of the police incident records and citizen survey responses using the web application. As this was the first time CoDs used web applications for real implementation and analyzed charts, they struggled with quickly reading and interpreting the data tables and charts, necessitating additional support from researchers. Although the CoD did not make any policy decisions, they used the dashboard to engage with residents and NGOs about the findings.

\section{Citizen Survey Questionnaires}\label{sec:citizen-surveys}

\subsection{Feelings of Insecurity}

The items in this survey were adapted from Gray et al. \cite{Emily2010Feelings}, with modifications made to fit the context of the present study.

\textit{To be filled by Surveyor}:
\begin{enumerate}
    \item Survey District
\item Survey Neighborhood
\end{enumerate}
\textit{For the Respondents}:\\
\textbf{Demographic information}
\begin{enumerate}
    \item Age Group: (\textit{18-29, 30-41, 42-53, 54-65, 66 and above})
\item Gender: (\textit{Woman, Man, Others, Prefer not to say})
\item Profession: (\textit{Student, Salaried worker, Seeking opportunities, or Self-Employed, Others})
\item Resident in the surveying neighborhood: (\textit{Yes/No})
\item If not, the residence neighborhood: 
\item How frequently do you visit this neighborhood (\textit{every day, once a week, once a month, two to three times a month, twice a year, once a year, prefer not to say})
\end{enumerate}
\textbf{Feelings of Insecurity}

\begin{enumerate}
\item In the past six months, how frequently have you worried about theft on the street? (\textit{n times})
\item If at least once, on a scale of 1-5, how fearful did you feel on the last occasion? (\textit{1 = not at all fearful and 5 = extremely fearful})
\item In the past six months, how frequently have you worried about intoxicated people on the street? (\textit{n times})
\item If at least once, on a scale of 1-5, how fearful did you feel on the last occasion? (\textit{1 = not at all fearful and 5 = extremely fearful})
\item In the past six months, how frequently have you worried about hooliganism, including petty on the street? (\textit{n times})
\item If at least once, on a scale of 1-5, how fearful did you feel on the last occasion? (\textit{1 = not at all fearful and 5 = extremely fearful})
\item In the past six months, how frequently have you worried about the people lying/sleeping on the street? (n times)
\item If at least once, on a scale of 1-5, how fearful did you feel on the last occasion? (\textit{1 = not at all fearful and 5 = extremely fearful})
\item On a scale of 1 to 5, how safe and secure do you feel in public transport after dark? (\textit{1 = not at all safe and 5 = completely safe; I never travel by public transport})
\item In the past six months, have you avoided certain streets or areas during the day, because of fear of crime?
\begin{itemize}
    \item If YES, provide the name of the most frequently avoided street during the day: 
\end{itemize}
\item In the past six months, have you avoided certain streets or areas during the night, because of fear of crime?
\begin{itemize}
    \item If YES, provide the name of the most frequently avoided street at night:
\end{itemize}
\item Please tick up to three locations that would benefit from police patrols. 
\begin{itemize}[label=$\square$]
    \item in your residential area
    \item at your public transport stop
    \item in the neighborhood centre
    \item at the shopping centre
    \item other (Please specify)
\end{itemize}

\end{enumerate}

\subsection{Residents' Cohesion and Disorder in the Neighborhood}

The items in this survey were adapted from Reid et al. \cite{Reid2020Perceptionsmodel}, with modifications made to fit the context of the present study.

\textit{To be filled by Surveyor}:
\begin{enumerate}
    \item Survey District
\item Survey Neighborhood

\end{enumerate}
\textit{For the Respondents}:\\
\textbf{Demographic Information}
\begin{enumerate}
    \item Age Group: (\textit{18-29, 30-41, 42-53, 54-65, 66 and above})
\item Gender: (\textit{Woman, Man, Others, Prefer not to say})
\item Profession: (\textit{Student, Salaried worker, Seeking opportunities, or Self-Employed, Others})
\end{enumerate}

\textbf{Cohesion and Social Order}
\begin{enumerate}
    \item In the past six months, on a scale of 1-5, do you feel fearful about the following problems in the area where you live: (\textit{1 = not at all fearful and 5 = extremely fearful})
    \begin{itemize}
\item Abandoned buildings
\item Littering/garbage in streets
\item People drinking or drunk in the street
\item Abandoned vehicles 
\item Traffic congestion
\item Vandalism
\end{itemize}
\item On a scale of 1-5, please state to what extent you agree or disagree with the following statements: (\textit{1 = strongly disagree and 5 = strongly agree})
\begin{itemize}
    \item People in the area where I live are willing to help their neighbors
    \item People in the area where I live know each other well
    \item If I sensed trouble in the area where I live, I could rely on the people who live there for help
    \item People in the area where I live can be relied upon to call the police if they see something suspicious
    \item People in this neighborhood can be trusted
\end{itemize}
    \item If you suspect a burglary in your street, how would you respond?
    \begin{itemize}[label=$\square$]
        \item call the police on 110
        \item call the police on 1188
        \item contact a neighbor
        \item at the shopping centre
        \item take no action
        \item other (Please specify)
    \end{itemize}
\end{enumerate}

\subsection{Perceived Risk of Harm and Trust in police}

The items in this survey were adapted from Gray et al. \cite{Emily2010Feelings}, and  Reid et al. \cite{Reid2020Perceptionsmodel}, with modifications made to fit the context of the present study.

\textit{To be filled by Surveyor}:
\begin{enumerate}
    \item Survey District
\item Survey Neighborhood
\end{enumerate}

\textit{For the Respondents}:\\
\textbf{Demographic Information}
\begin{enumerate}
    \item Age Group: (\textit{18-29, 30-41, 42-53, 54-65, 66 and above})
\item Gender: (\textit{Woman, Man, Others, Prefer not to say})
\item Profession: (\textit{Student, Salaried, Seeking opportunities, or Self-Employed, Others})
\end{enumerate}

\textbf{Perceived risk of harm and Trust in police}
\begin{enumerate}
    \item On a scale of 1-5, please state to what extent you agree or disagree with the following statements: (\textit{1 = strongly disagree and 5 = strongly agree})
        \begin{itemize}
            \item Our local police react promptly if called in an emergency
            \item Our local police deal effectively with crime issues in the area where I live
            \item Our local police treat everyone fairly
            \item Our local police listen to people's concerns
            \item Our local police are dealing with the things that matter in the area where I live
            \item Our local police do a good job in the area where I live
        \end{itemize}
        
    \item On a scale of 1-5, how likely do you think each of the following crimes is to happen during the next six months: (\textit{1 = not at all likely and 5 = extremely likely})
        \begin{itemize}
            \item Someone breaking into your home
            \item Someone stealing items that belong to you without using force 
            \item Someone taking something from you by force or threat of force
            \item Someone harassing, threatening, or verbally abusing you
            \item Someone beating or attacking you
        \end{itemize}

    \item In the past 3 months, how often were you worried about: (\textit{n times}) 
        \begin{itemize}
            \item Someone breaking into your home
            \item Someone stealing items that belong to you without using force
            \item Someone taking something from you by force or threat of force
            \item Someone harassing, threatening or verbally abusing you
            \item Someone beating or attacking you
            \item Someone threatening or attacking family members
            \item Someone threatening, or attacking your friends
        \end{itemize}
\end{enumerate}

\section{Surveyors' Feedback Questionnaire}\label{sec:surveyors-feedback}
To improve the survey and the surveying experience of the civic engagement tool, the CoDs collect feedback from the patrol officers, district coordinators, and volunteers, following this questionnaire.
\textbf{Survey Information}
\begin{enumerate}
    \item Surveyed District
    \item Surveyed Neighborhood
    \item Survey Period
\end{enumerate}

\textbf{Demographic Information}
\begin{enumerate}
    \item Age Group: (\textit{18-29, 30-41, 42-53, 54-65, 66 and above})
    \item Gender: (\textit{Woman, Man, Others, Prefer not to say})
    \item Category: (\textit{Patrol officer, district coordinator, NGO volunteer})
\end{enumerate}

\textbf{Surveying}
\begin{enumerate}
    \item On a scale of 1--5, how would you rate your overall experience of the survey? (\textit{1 = Terrible, 2 = Poor, 3 = Neutral, 4 = Good, and 5 = Excellent})
    \item On a scale of 1--5, how enthusiastic/willing were the respondents about the survey: (\textit{1 = not at all enthusiastic/willing and 5 = extremely enthusiastic/willing})
    \item On a scale of 1--5, how difficult did you feel about the following items: (\textit{1 = not at all difficult and 5 = extremely difficult})
    \begin{itemize}
        \item Explaining the survey to the respondents
        \item Identifying the respondents
        \item Identifying the respondents with age diversity
        \item Identifying the respondents with gender diversity
        \item Identifying the respondents who are residents of the same neighborhood
    \end{itemize}
    \item Will you be interested in continuing as the surveyor for the next edition in this neighborhood?: (\textit{Yes, No, Maybe})
    \item Were there any questions that the citizens did not prefer to answer?
    \item What was your favorite part of the surveying?
    \item Could you tell us about your least favorite part of the surveying?
    \item Do you have any suggestions for the next edition of the survey?
    \item Do you have any suggestions/comments you received from the citizens during the survey to share with us?
\end{enumerate}

\section{Design Assessment}\label{app:toolcomp-evaluation}

\begin{figure}[ht]
  \centering
  \begin{subfigure}{0.48\textwidth}
      \includegraphics[width=1\linewidth]{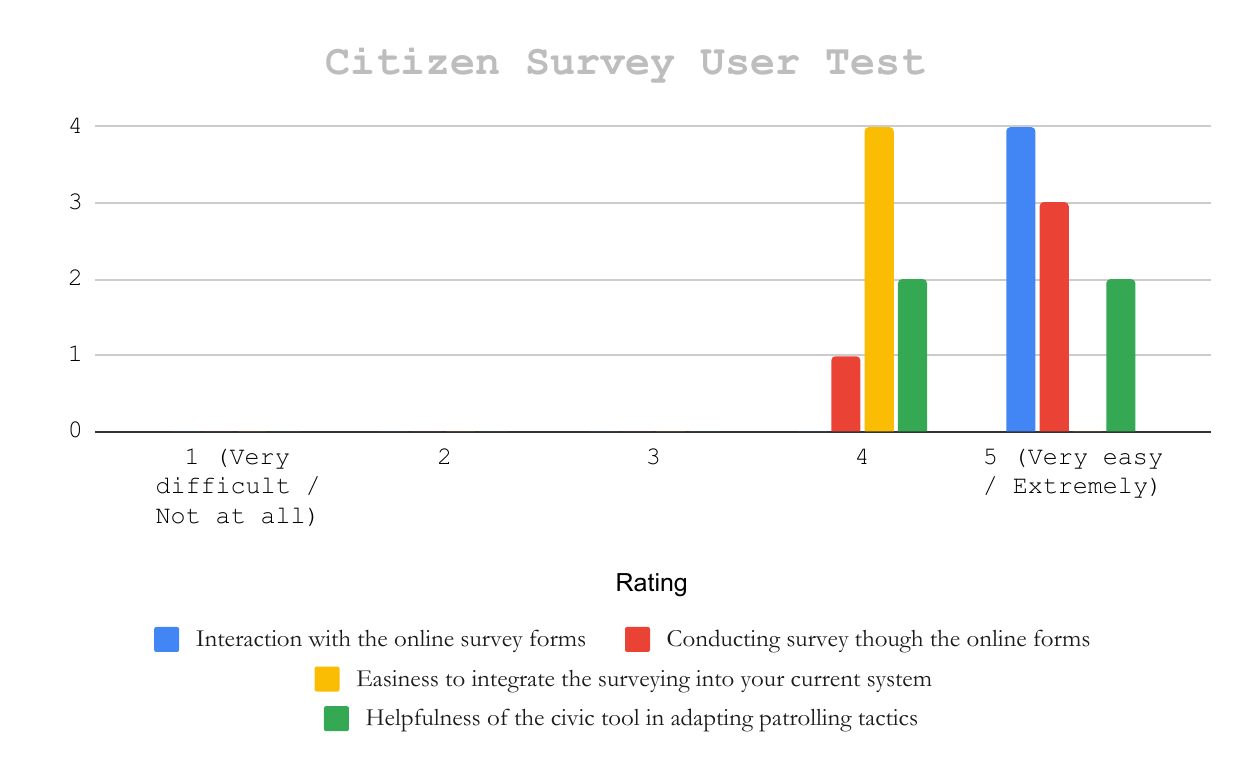}
      \subcaption{Surveyors Feedback on Citizen Survey Tool Component (4 respondents).}
      \Description{Bar chart titled ``Citizen Survey User Test'' summarizing ratings from four participants on four aspects of the surveying component: interaction with the online survey form, conducting the survey through the online form, ease of integrating survey data into the current system, and perceived helpfulness of the tool for adapting patrolling tactics. Bars show mostly high ratings (around 3–4) across items, with slightly lower ratings (around 1–2) on some items for one or two participants.}
      \label{fig:surveyor-tool-eval}
  \end{subfigure}
\hfill
  \begin{subfigure}{0.48\textwidth}
      \includegraphics[width=1\linewidth]{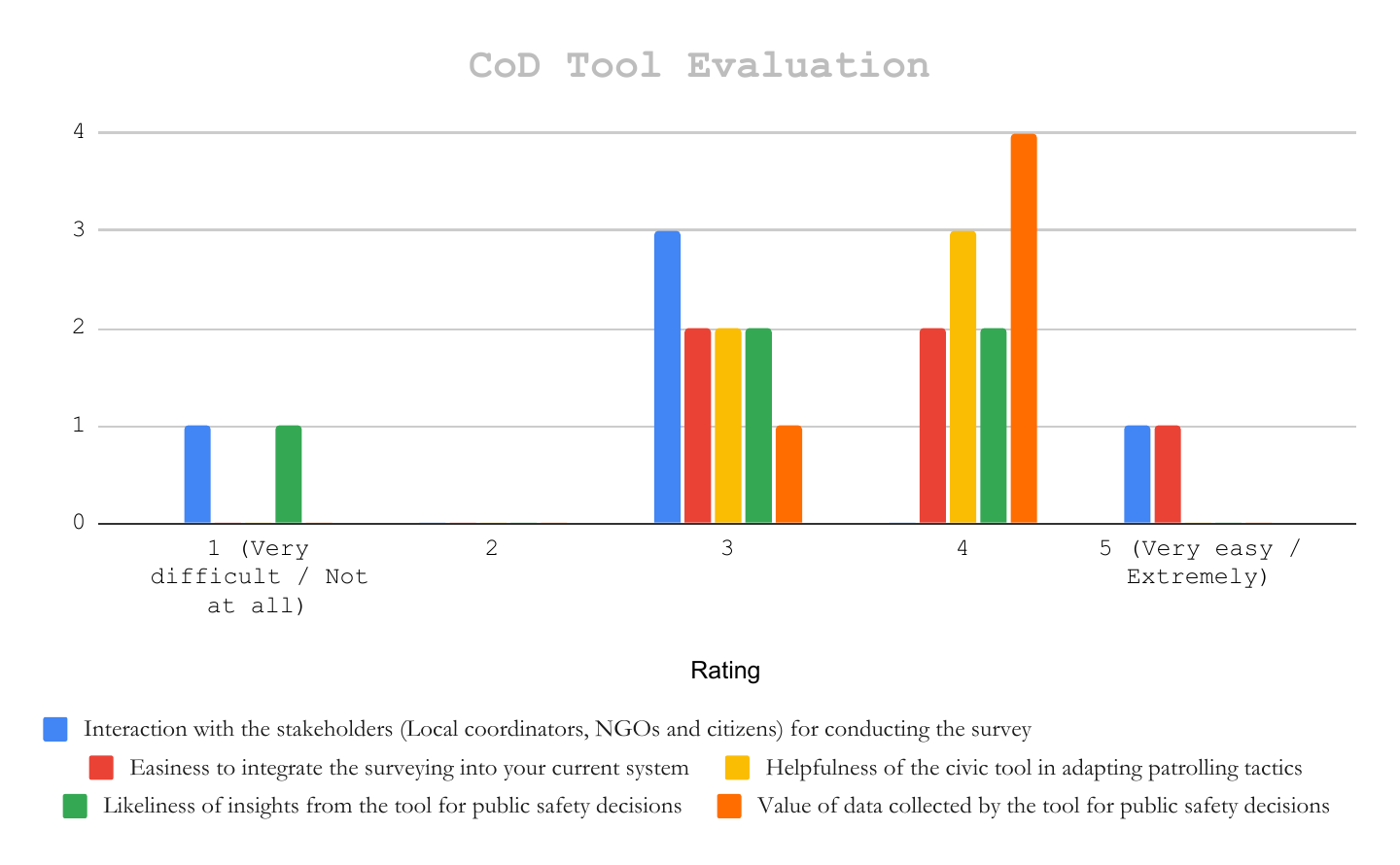}
      \subcaption{Chiefs of Districts Feedback on Tool Components (5 respondents).}
      \Description{Bar chart titled ``CoD Tool Evaluation'' showing distributions of five items rated on a 1–5 scale (1 = very difficult/not at all; 5 = very easy/extremely), with y-axis as number of responses (N=5). Items: interaction with stakeholders for conducting the survey (ratings: 1x1, 3x3, 5x1); ease of integrating surveying into the current system (3x2, 4x2, 5x1); helpfulness for adapting patrolling tactics (3x2, 4x3); likeliness that the tool produces insights for public safety decisions (1x1, 3x2, 4x2); and value of data collected for public safety decisions (3x1, 4x4). Overall, responses cluster at 3–4, with few low (1) and some top (5) ratings.}
      \label{fig:cod-tool-eval}
  \end{subfigure}

  \caption{Responses to Tool Components Evaluation}
  \label{fig:tool-evaluation}
\end{figure}

\section{List of Participants} \label{app:participants}

\begin{table}[H]
\centering
  \caption{Participants in the Preparatory Workshop} \label{tab:participants-preparatory}
  \Description{Table listing seven preparatory workshop participants (PW1--PW7) with their organization and role, including municipal police specialists, civil-society evaluation specialists, a Belgian municipal police prevention/security lead, and two academic researchers; sex is indicated for each participant.}

  \begin{tabular}{p{0.07\linewidth}p{0.35\linewidth}p{0.42\linewidth}}
    \toprule
    \textbf{No.} & \textbf{Organization} & \textbf{Role (Sex)} \\
    \midrule
    PW1 & Riga municipal police & Chief Specialist (M) \\
    PW2 & Riga municipal police & Chief Specialist (M) \\
    PW3 & Public/Civil Society Evaluation & Research and Evaluation Specialist (F)  \\
    PW4 & Public/Civil Society Evaluation & Research and Evaluation Specialist (M) \\
    PW5 & A Belgian city's\newline municipal police & Head of the Prevention and Security (M) \\
    PW6 & University & Human-Centered Design Researcher (M) \\ 
    PW7 & Research Institute & Social Computing\newline Researcher (F) \\ \bottomrule
\end{tabular}
\end{table}

\begin{table}[H]
\centering
  \caption{Participants in the Citizen Survey User Testing} \label{tab:cit-survey}
  \Description{Table listing four participants (CS1--CS4) in citizen survey user testing with organization and role: two patrol officers from the municipal police and two district coordinators from the municipal administration; sex is indicated for each participant.}

  \begin{tabular}{p{0.1\linewidth}p{0.35\linewidth}p{0.4\linewidth}}
    \toprule
    \textbf{No.}  & \textbf{Organization} & \textbf{Role (Sex)} \\
    \midrule
    CS1 & Riga municipal police  & Patrol Officer (F) \\
    CS2 & Riga municipal police  & Patrol Officer (M) \\
    CS3 & Riga municipal\newline administration & District Coordinator (M)  \\
    CS4 & Riga municipal\newline administration & District Coordinator (M) \\
\bottomrule
\end{tabular}
\end{table}

\begin{table}[H]
\centering
  \caption{Participants in the Web Application User Testing} \label{tab:web-app}
  \Description{Table listing five participants (WA1--WA5) in web application user testing, all from the municipal police, with roles in district leadership and public order; sex and age are indicated for each participant.}

  \begin{tabular}{p{0.1\linewidth}p{0.35\linewidth}p{0.42\linewidth}}
    \toprule
    \textbf{No.} & \textbf{Organization} & \textbf{Role (Sex/Age)} \\
    \midrule
    WA1 & Riga municipal police & Chief Specialist (F/46) \\
    WA2 & Riga municipal police & Chief of Central\newline District (M/39) \\
    WA3 & Riga municipal police & Deputy Chief of Central\newline District (M/44) \\
    WA4 & Riga municipal police & Chief Inspector from\newline Central District (M/36) \\
    WA5 & Riga municipal police & Head of the Public\newline Order Protection of\newline Central District (M/38) \\
\bottomrule
\end{tabular}
\end{table}

\begin{table}[H]
\centering
  \caption{Participants in the Continuous Assessment of the Tool} \label{tab:participants-continious}
  \Description{Table listing eight participants (CA1--CA8) in continuous assessment with organization and role: municipal police specialists and district leadership, a deputy mayor and district coordinator, a project coordinator from a European urban safety network, a research institute member, and a Swedish municipal coordinator; sex is indicated for each participant.}

  \begin{tabular}{p{0.1\linewidth}p{0.35\linewidth}p{0.42\linewidth}}
    \toprule
    \textbf{No.}  & \textbf{Organization} & \textbf{Role (Sex)} \\
    \midrule
    CA1 & Riga municipal police & Chief Specialist (M) \\
    CA2 & Riga municipal police & Chief Specialist (M) \\
    CA3 & Riga municipal police & Chief of Central District (M) \\
    CA4 & Riga municipal\newline administration & Deputy Mayor (F) \\
    CA5 & Riga municipal\newline administration & District Coordinator (M) \\
    CA6 & European network\newline focused on\newline urban safety & Project Coordinator (F) \\
    CA7 & Research Institute & Social Computing\newline Researcher (M) \\ 
    CA8 & Swedish city municipal administration & Security and Radicalization Prevention Coordinator (F) \\ \bottomrule
\end{tabular}
\end{table}

\end{document}